\pdfoutput=1
\documentclass[5p,10pt,number]{elsarticle}
\usepackage[english]{babel}
\usepackage{graphicx}
\usepackage[utf8]{inputenc}
\usepackage[T1]{fontenc}
\usepackage{url}
\usepackage{natbib}
\usepackage{longtable}
\usepackage{multirow}
\usepackage{colortbl}
\usepackage{amsmath,amssymb}
\usepackage[colorlinks,citecolor=blue,filecolor=blue,linkcolor=blue,urlcolor=blue]{hyperref}
\newcommand{\e}{{\rm e}}
\newcommand{\figdir}{.}

\begin{document}

\title{Transmission grid extensions during the build-up of a fully renewable pan-European electricity supply} 

\author[fias]{S.~Becker\corref{cor}}
\ead{becker@fias.uni-frankfurt.de}

\author[au]{R.\,A.~Rodriguez}

\author[au]{G.\,B.~Andresen}

\author[fias]{S.~Schramm}

\author[au]{M.~Greiner}

\cortext[cor]{Corresponding author}

\address[fias]{
  Frankfurt Institute for Advanced Studies,
  Goethe-Universit\"at, 
  60438~Frankfurt am Main, Germany
  }
\address[au]{
  Department of Engineering and Department of Mathematics,
  Aarhus University, 
  8000~Aarhus~C, Denmark
  }

\begin{abstract}
Spatio-temporal generation patterns for wind and solar photovoltaic power in
Europe are used to investigate the future rise in transmission needs with an
increasing penetration of the variable renewable energy sources (VRES) on the
pan-European electricity system. VRES growth predictions according to the
official National Renewable Energy Action Plans of the EU countries are used and
extrapolated logistically up to a fully VRES-supplied power system. We find that
keeping today's international net transfer capacities (NTCs) fixed over the next
forty years reduces the final need for backup energy by 13\,\% when compared to
the situation with no NTCs. An overall doubling of today's NTCs will lead to a
26\,\% reduction, and an overall quadrupling to a 33\,\% reduction. The
remaining need for backup energy is due to correlations in the generation
patterns, and cannot be further reduced by transmission. The main investments in
transmission lines are due during the ramp-up of VRES from 15\,\% (as planned
for 2020) to 80\,\%. Additionally, our results show how the optimal mix between
wind and solar energy shifts from about 70\,\% to 80\,\% wind share as the
transmission grid is enhanced. Finally, we exemplify how reinforced transmission
affects the import and export opportunities of single countries during the VRES
ramp-up.  
\end{abstract}


\maketitle

\section{Introduction}

In order to reach the 2020 and 2050 CO$_2$ reduction goals of the European
Union, a high share of renewable electricity generation from weather-dependent
sources, mainly wind and solar PV power, is inevitable \cite{ecf2050}. According
to \cite{Jacobson:2009nx}, this is both an economically and environmentally
viable solution. As opposed to traditional electricity generation from
dispatchable power plants, these variable renewable energy sources (VRES) are
intermittent. Desirable features of an electricity system with high VRES shares
are low needs of storage, transmission and (conventional) back\-up power, but
also minimal excess generation of VRES. Based on the mismatch between
weather-determined generation data and historical load, optimal mixes of wind
and solar energy with respect to different objectives have been derived
\citep{heide2010,heide2011,schaber2}. Moreover, storage needs on different time
scales have been looked at and strong synergies between short-term
storage, long-term storage and backup power generation have been discovered
\citep{morten}. Most recently, the transmission needs of a fully VRES-supplied
Europe have been investigated and different grid extension scenarios in order to
minimise the need for backup energy were examined \citep{rolando}. In the spirit
of this work, we now proceed to look at not only the fully renewable end-point
scenario, but also at possible pathways to it. 

As opposed to other grid extension studies such as
\cite{schaber2,schaber1,energynautics,steinkegrid}, we do not follow an overall
cost-optimal approach with a simplified transmission paradigm, but focus on
\emph{maximum usage and optimal sharing of VRES} at a \emph{minimal
transmission} capacity layout. Our results are independent of cost assumptions. 
At a later stage, our approach can be extended to include an economical evaluation 
of how a targeted reduction in backup energy can be achieved at a minimal 
transmission investment. We use a straight-forward generalization of the
physical DC power flow paradigm together with the most localised equal-time
matching of excess-power and deficit-power regions. This yields a
lower bound on the total necessary link capacity. It has to be noted that
market-driven power transfer will in general lead to more flow. We also do not
take the necessary upgrades of the country-internal grids into account. We
calculate how much inter-country transmission capacity is needed to reduce the
necessary total backup energy by a certain percentage. 

We strive to answer the following questions: Which pan-European transmission
needs do arise where, and when? How can transmission mitigate backup needs? What
are other benefits of reinforced transmission grids, e.g.\ facilitated trade, and
what investments are required? The latter two questions have already been
addressed in \cite{rolando} for a fully renewable end-point scenario. Here, we
extend this discussion to the transitional pathways.

The paper is organised in the following way: In Sec.\ 2, we describe the load
and VRES generation data used in this work, the assumptions made for the growth
of VRES installation, and the power flow calculations. Sec.\ 3 presents results
on the time-dependent reinforcements of the transmission grid during the VRES
ramp-up necessary to reduce backup energy by a given amount. It also includes a
discussion of the impact of transmission on the optimal mix of wind and solar
energy and on the future import and export opportunities of single countries.
Sec.\ 4 concludes the paper.

\section{Data and methodology}

\subsection{Weather, generation and load data}

The weather data set used covers the eight year period 2000-2007 with a temporal
resolution of one hour and a spatial resolution of $\rm 47\,km\times 47\,km$.
Encompassed are the European Union as well as Norway, Switzerland, and the
Balkans. The weather data were converted into wind and solar PV power generation
time series for all grid points as described in \cite{heide2010}. These were
further aggregated to country level, ignoring any national transmission
bottlenecks.

In addition to weather data, historical load data with hourly resolution were
obtained for all countries. They were either downloaded directly from UCTE (now
ENTSO-E) \cite{entsoe_load} for the same eight year period, or extrapolated from
the UCTE data for countries where load data were not available throughout the
eight year simulation period. For additional details see \cite{iset2008}.
Finally, the load time series of each country was detrended from an average
yearly growth of about 2\,\%.

Data for the Baltic have been synthesised based on corresponding time series for
Finland and Poland since they were not included in the original data set.
 
By combining load and (scaled) VRES power generation, we calculate their hourly
mismatch for each country as expressed by Eq.\ \eqref{eq:mism} below. When the
mismatch is positive, VRES generation is in surplus, i.e.\ it exceeds the load,
and when it is negative, a deficit of VRES generation occurs as compared to the
load. 
\begin{align}
  \Delta_n(t)  =  \gamma_n \langle L_n\rangle 
                         \left(  
                         \alpha^{\rm W}_n \frac{G^{\rm W}_n(t)}{\langle G^{\rm W}_n \rangle} 
                         + (1-\alpha^{\rm W}_n) \frac{G^{\rm S}_n(t)}{\langle G^{\rm S}_n \rangle}
                         \right)
                         - L_n(t) \,,
\label{eq:mism}
\end{align}
where $L_n$ is the load at node (country) $n$, $G^{\rm W}_n$  denotes the
corresponding wind and $G^{\rm S}_n$ the solar PV generation time series. The
node $n$'s VRES \emph{penetration}, i.e.\ the ratio between mean VRES generation
and load, is denoted $\gamma_n$. $0\leq \alpha^{\rm W}_n\leq 1$ is the share of
wind in VRES at node $n$; we also refer to it as the relative \emph{mix} of wind
and solar energy. $(1-\alpha^{\rm W}_n)$ is the corresponding share of solar PV,
and $\langle .\rangle $ denotes the time average of a quantity.

The negative part of a country's mismatch may partly be covered by imports from
other countries. What is still missing after imports is what has to be covered
by the local dispatchable backup system. We term it \emph{balancing}, as it is
required to maintain balance between supply and demand in the power system.
Here, every form of electricity generation other than VRES is subsumed under
balancing.

\subsection{Growth of VRES 1990-2050}
\label{sec:logfit}

\subsubsection*{Overview}
In order to model the growth of VRES installation from today's values up to a
fully VRES-supplied energy system, we let $\alpha^{\rm W}_n$ and $\gamma_n$ from
Eq.\ \eqref{eq:mism} depend smoothly on a reference year. The reference years
correspond to real years in the sense that historical penetrations of wind and
solar power are made to follow historical values. In a similar fashion, future
penetrations are based on official 2020 targets and 2050 assumptions. $\gamma_n$
and $\alpha_n^{\rm W}$ are obtained by fitting growth curves to historical and
targeted penetrations. The year variable of the fit is termed reference year to
emphasise that the fit does not exactly pass through neither the historical nor
the targeted values.

\subsubsection*{Historical data and 2020 targets} The historical wind and solar
penetrations originate from Eurostat \citep{eurostat} for EU member states as
well as Switzerland, Norway, and Croatia, and from the IEA \citep{ieabalkan} for
the other Balkan countries. 

The 2020 targets for EU member states are taken from their official National
Renewable Energy Action Plans \cite{nreap}. In the case of Denmark, this target
has already been revised because of the strong growth in wind installations, and
we consequently use the new target \cite{DK_new}. For Switzerland, the Energy
Strategy 2050 of the Swiss government and the corresponding scenario from a
consulting firm is employed \cite{CHbundesrat,prognos}. For Croatia, we use the
Croatian energy strategy as officially communicated in \cite{croatia_2020}.
These figures are also applied to the other Balkan states, since no other data
source has been found. For Norway, the 2020 targets are estimates of the
independent research organisation SINTEF \cite{sintef}.

\subsubsection*{2050 targets}
For the reference year 2050, we assume a very ambitious end-point scenario by
setting the target penetration of VRES to 100\,\% of the average electricity
demand for all countries ($\gamma_n = 1$). However, even at this penetration, a
backup system of dispatchable power plants is needed to ensure security of
supply when the production from VRES does not meet the demand. The minimum
balancing energy that must be provided by the backup system was investigated in
\cite{heide2011,morten,rolando}, and for a penetration of 100\,\%, it amounts on
average to between 15\,\% and 24\,\% of the demand, depending on the strength of
the transmission grid. In a fully renewable power system, this energy must be
provided by dispatchable renewable technologies such as hydro power and biomass,
or from re-dispatch of earlier stored VRES-surplus. In general, conventional
fossil and nuclear plants can also be used. 

The official goal of the European Union is to reduce CO$_2$ emissions by 80\,\%
before 2050 \cite{eu2050}. It is argued in \cite{ecf2050} that to reach this
goal it will be necessary to decarbonise the electricity sector almost
completely. The ambitious target of a VRES penetration of 100\,\% ($\gamma_n=1$)
by 2050 is consistent with this goal as the required balancing energy could be
provided by a combination of dispatchable renewable resources such as hydro
power and biomass, possibly in combination with storage as investigated in
\cite{morten}. More conservative end-point scenarios with a lower VRES
penetration, e.g.\ those of \cite{ecf2050,energynautics}, can easily be
encompassed implicitly by shifting the 100\,\% VRES targets to later times,
e.g.\ 2075 or 2100.

It is reasonable to be restrictive in the assumptions on the contribution from
non-variable RES, which are mainly biomass and hydro power. As summarised in
Tab.\ \ref{tab:rpot}, growth is severely constrained for both. For biomass, this
is firstly because agricultural areas not needed for food production are limited
and secondly because it is also commonly used for bio-fuel and heating. As a
result, its contribution to electricity generation is only expected to double
during the period 2010-2020, until it covers about 7\,\% of the total (2007)
load. In sharp contrast, growth by factors of four to five for the VRES
technologies are expected. Concerning hydro power, most of what is feasible is
already in use today. So, further substantial growth is not expected in the EU,
cf.\ also \cite{Lehner:2005cr}. The expected hydro installation in 2020 in the EU
is able to cover 11\,\% of the 2007 load (Tab.\ \ref{tab:rpot}); note that the
significant resources of Norway are not included here, which would yield another
4\,\%.  We also do not include explicitly other forms of renewable energy, such
as tidal, ocean, wave or geothermal energy, because these are still in early
stages of their development and whether or not they will yield a substantial
contribution at some point remains uncertain, cf.\ Tab.\ \ref{tab:rpot}.

\begin{table}[!t]
\caption{
Planned development of different RES technologies in the EU-27: Historical and
projected gross electricity generation. Although the EU-27 is not identical to
our simulated region, the deviations are small and therefore, the EU projections
can serve as a proxy to the economical renewable potentials we are interested
in. Data taken from the National Renewable Energy Action Plans \cite{nreap}, as
compiled by the Energy Research Centre of the Netherlands \cite{ecn}.  
}
\label{tab:rpot}
\centering
\begin{tabular}{l r r}
\hline
Electricity generation & \multicolumn{2}{c}{Year}\\
(TWh/yr) & 2010 & 2020 \\
\hline
Wind (on- and offshore) & 165 & 495 \\
Solar (PV and CSP) & 21 & 103 \\
Hydro & 343 & 369 \\
Biomass & 114 & 232 \\
Geothermal & 6 & 11 \\
Ocean (heat, wave, and tidal) & 1 & 7 \\
\hline
\end{tabular}
\end{table}

Nuclear power and some other conventional sources of electricity, e.g.\ fossil
fuel plants with carbon capture and storage (CCS) technologies, could also
provide CO$_2$-free balancing power or replace some or all VRES. However, the
ever-growing acceptance problems of nuclear power makes the future role of this
technology uncertain. Such concerns have already led to accomplished or planned
phase-outs or bans in several countries, such as Ireland, Italy, Austria,
Denmark, Belgium, Germany, and Switzerland. For similar reasons, we also rule
out CCS. Fusion power could become an option in the far future, but it will not
be commercially available before 2040 or later \cite{iter}, and consequently it
cannot play a role in the transition to a decarbonised electricity system Europe
is facing before 2050.

In summary, this means that most of the growth of CO$_2$-free electricity
generation until 2050 has to come from wind and solar power.

The mix between wind and solar power in 2050 is chosen such that the balancing
energy becomes minimal for each single country for the base scenario. As an
example, Fig.\ \ref{fig:logfit_de}c shows the balancing energy as a function of
the mix for Germany, for a penetration $\gamma_{\rm DE}=1.0$, with a clear
minimum at 0.72. For all other countries a similar behaviour can be observed,
and the average mix of the base scenario becomes 0.71, with individual countries
ranging from 0.64 for Croatia to 0.85 for Norway \cite{rolando}.

\subsubsection*{Fit function}
The historical and targeted wind and solar penetrations are fitted with logistic
growth curves. Logistic curves have proven to be able to successfully model the
diffusion of new technologies across various fields such as infrastructures,
e.g.\ canals, railroads, roads \cite{infrastructures}, electrification, and
household appliances such as refrigerators and dishwashers \cite{greatcentury}.
The case of energy transitions is discussed in \cite{gruebler}, and the current
switch to CO$_2$-neutral generation is further analysed in \cite{wilson}. We
assume that the penetration of wind and solar PV electricity production will
follow a similar growth for each individual country.

A general logistic function is given by
\begin{align}
  f(y;y_0,a,b,m)
    =  \frac{a\cdot b \cdot \e^{m(y-y_0)}}{a(\e^{m(y-y_0)}-1)+b} \, .
\label{eq:logfit}
\end{align}
In our application, $f$ denotes either the wind ($\gamma_n \cdot \alpha^{\rm
W}_n)$ or the solar $(\gamma_n\cdot (1-\alpha^{\rm W}_n))$ penetration. The
reference year is denoted $y$, with $y_0$ and $a,b,m\geq 0$ being the fit
parameters. $a$ is the value of $f$ in year $y_0$, $b$ is the limiting value for
late years, and $m$ is a proxy to the maximal slope. This function is
least-square fitted to historical and projected wind or solar penetration data.

\begin{figure*}[!ht]
\begin{center}
\includegraphics[width=0.370\textwidth]{\figdir/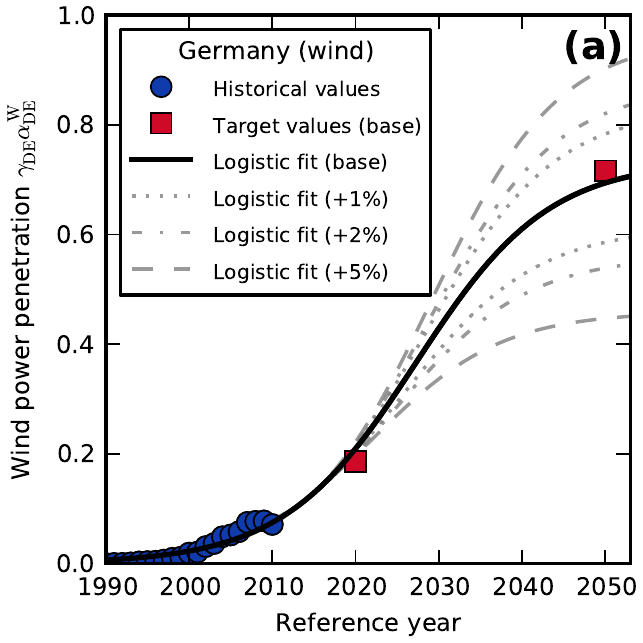}
\includegraphics[width=0.370\textwidth]{\figdir/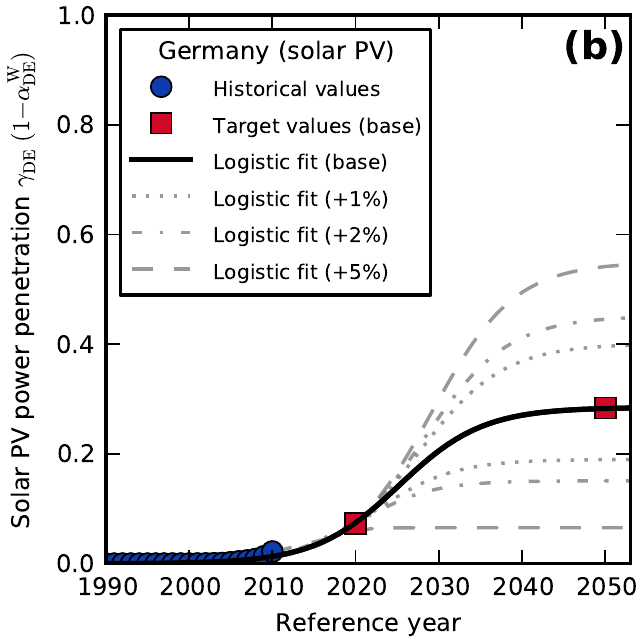}
\includegraphics[width=0.245\textwidth,type=pdf,ext=.pdf,read=.pdf]{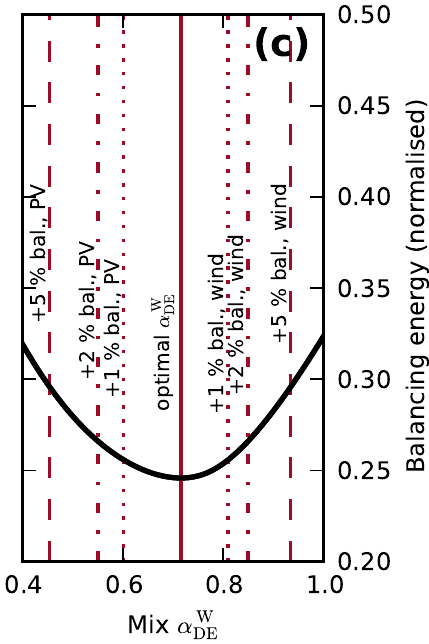}
\caption{
Logistic fits to historical and targeted wind (a) and solar PV (b) penetration
in Germany 1990 - 2050. In panel (c), balancing energy as a function of the
wind/solar mix for Germany is shown, for a renewable penetration of $\gamma_{\rm
DE}=1.0$. The solid vertical line in this panel points out the balancing optimal
mix, and the dotted, dashed-dotted and dashed lines indicate the mixes that lead
to balancing needs increased by 1\,\%, 2\,\%, and 5\,\% of the load,
respectively. These mixes are used for seven different 2050 targets in panels
(a) and (b): the optimal mix corresponds to the base scenario (solid black
lines), and the other mixes lead to the three wind heavy and three solar heavy
logistic growth scenarios (grey broken lines). In calculating the balancing
needs of single countries as in (c), imports and exports are not considered.
Balancing is normalised by the average load.
}
\label{fig:logfit_de}
  \end{center}
\end{figure*}

\begin{figure*}[!ht]
\begin{center}
\includegraphics[width=0.49\textwidth]{\figdir/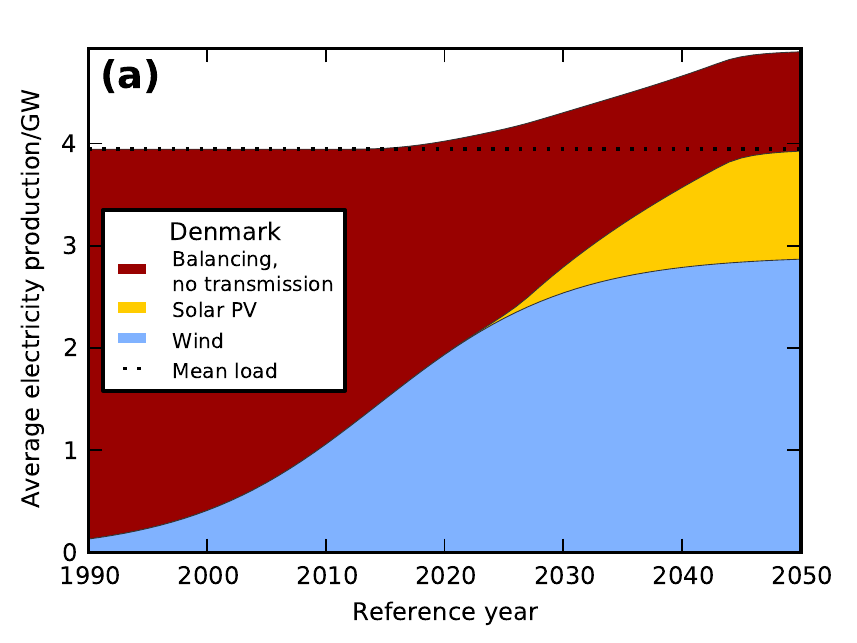}
\includegraphics[width=0.49\textwidth]{\figdir/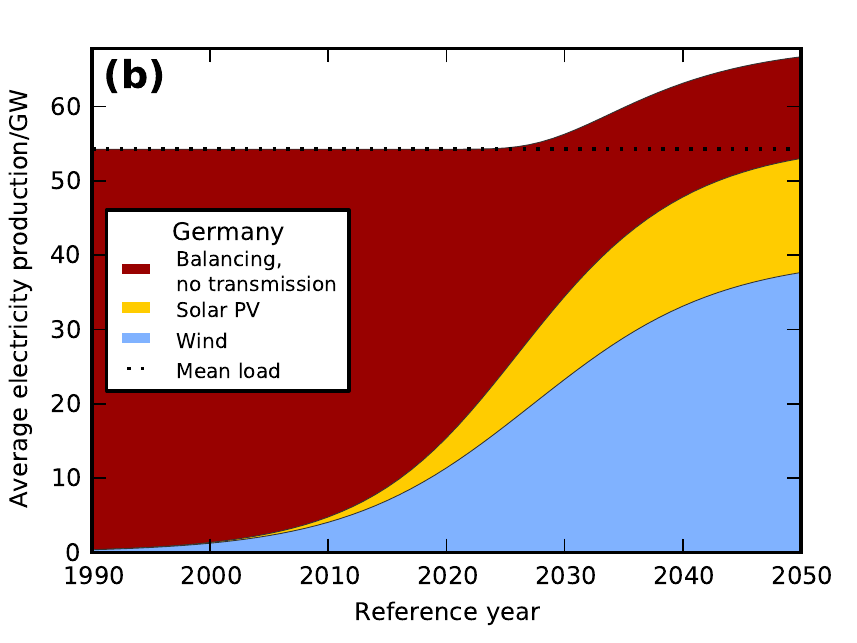}
\includegraphics[width=0.49\textwidth]{\figdir/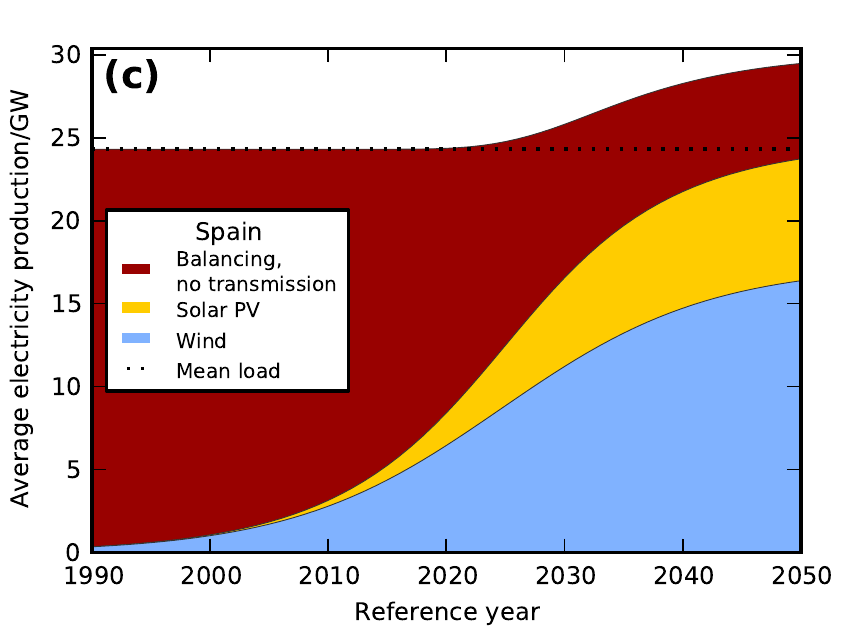}
\includegraphics[width=0.49\textwidth]{\figdir/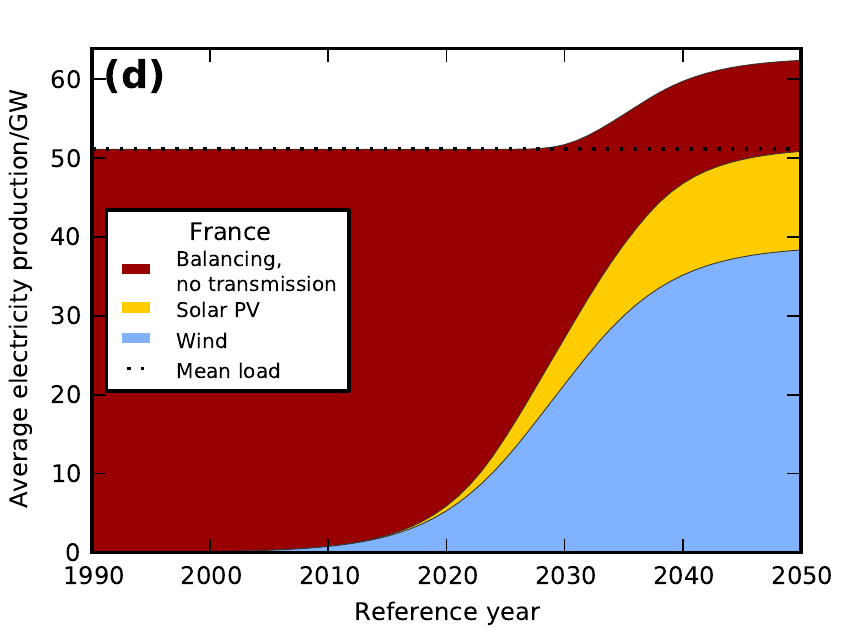}
\caption{
Logistic growth of wind (blue) and solar PV (yellow) power production, replacing
conventional balancing, for (a) Denmark, (b) Germany, (c) Spain, and (d) France.
Shown is the base scenario, where the single country balancing optimal mix is
reached in 2050. The balancing energy (red) has been determined with the
no-transmission layout. The conversion from the shares $\gamma_n \cdot
\alpha^{\rm W}_n$ (wind) resp. $\gamma_n\cdot (1-\alpha^{\rm W}_n)$ (solar PV)
to average generation in $\rm GW$ is done using the 2007 load.
}
\label{fig:logfits}
\end{center}
\end{figure*}

The logistic function is symmetrical to its inflection point. Since we target
relatively high end-point shares for 2050, this may lead to almost step-like
growth for countries which do not have a significant share of VRES yet. To
remove this artifact, we limit the growth rate to the rate necessary to replace
old production capacity for wind turbines and solar panels at the end of their
lifetime in the end scenario, i.e.\ we modify the fit function \eqref{eq:logfit}
by imposing a maximal slope. As a rough estimate, the lifetime is set to 20
years for both wind and solar installations for this purpose \cite{wslifetimes}.
Examples of these logistic fits can be seen in Fig.\ \ref{fig:logfit_de}, see
also Fig.\ \ref{fig:logfits}. Detailed numerical values are found in Tabs.
\ref{tab:endmix} and \ref{tab:logfit}. 

\subsubsection*{Alternative end-point mixes}
To investigate whether the optimal mix changes as soon as transmission comes
into play, we calculate logistic growth curves for a further range of end-point
mixes $\alpha_{\rm W}$ of wind and solar energy. The final mix is varied over a
range from roughly $\alpha^{\rm W}=0.40$ to $\alpha^{\rm W}=0.90$ wind share in
six additional scenarios apart from the optimal mix base scenario: three solar
heavy and three wind heavy scenarios, see Fig.\ \ref{fig:logfit_de}c. In the
base scenario, the end-point mix of each country is chosen as the mix that
minimises the average balancing energy of the country on its own, i.e.\ without
transmission. The wind heavy scenarios are defined by identifying the mixes to
the right of the minimum that lead to increases in balancing by 1\,\%, 2\,\%,
and 5\,\% of the average load. The three solar heavy scenarios are defined in
analogy to the left of the minimum. All seven scenarios are indicated in Fig.\ 
\ref{fig:logfit_de}c for Germany. Similar pictures emerge for the other
countries; the different end-point mixes are given in Tab.\ \ref{tab:endmix} In
the following, we examine the interplay between balancing and transmission in
all seven scenarios with a focus on the base scenario, and then investigate the
effect of transmission on the optimal mix comparing all scenarios.

\subsection{Power transmission and balancing}

The power transmission optimisation used in this paper applies a non-standard
approach designed to maximise the utilisation of VRES while minimising the total
need for transmission capacity. The optimisation is performed in two steps: The
total balancing $B_{\rm tot}$ is minimised first. Secondly, the dissipation in
the transmission network is minimised with $B_{\rm tot}$ constrained to its
minimum value. Here, the two steps are described in short. Additional details
can be found in \cite{rolando}. 

The physical DC approximation to the full AC power flow, see e.g.\ \cite{ooe},
is used in order to calculate the balancing need at each node and the flow
across each link. We conveniently assume equal susceptances for all the lines.
In this case, the net export $P_n$ of each of the $N$ nodes can be expressed in
terms of the $L$ flows $F_l$ via
\begin{align}
  P_n  =  \sum_{l=1}^L K_{nl}F_l.
\end{align}
Here, $K$ is the network's incidence matrix, i.e.\ 
\begin{align}
  K_{nl}  =  \begin{cases}
                 \phantom{-} 1 & \text{ if link $l$ starts at node $n$} \\
                 -1 & \text{ if link $l$ ends at node $n$} \\
                 \phantom{-} 0 & \text{ otherwise}
                 \end{cases}
\end{align}
It is not difficult to generalise this to the case of non-uniform susceptances;
these would appear in the entries of $K$. This formulation allows us to work in
terms of the flows alone, since balancing can now be expressed
as:\footnote{$(x)_-=\max\{-x,0\}$ denotes the negative part of a quantity $x$.}
\begin{align}
  B_{\rm tot}  =  \sum_{n=1}^N(\Delta_n-(K\cdot F)_n)_-=\sum_{n=1}^N B_n\,.
\end{align}
The balancing need $B_n$ at node $n$ is what is potentially left of a negative
mismatch $\Delta_n$ after it has been reduced by the net imports $(K\cdot F)_n$.

If the line capacities are constrained by (possibly direction dependent) net
transfer capacities (NTCs), $h_{-l}\leq F_l\leq h_l$, these constraints have to
be included. We end up with two minimisation steps, the first representing
maximal sharing of renewables, and the second minimizes transmission:
\begin{equation}
\begin{split}
  \text{Step 1: }&
    \min_{h_{-l}\leq F_l \leq h_l}  \sum_{i=1}^N(\Delta_n-(K\cdot F)_n)_- = B_{\rm min}\\
  \text{Step 2: }&
    \min_{\substack{h_{-l}\leq F_l \leq h_l \\ \sum_{i=1}^N (\Delta_n-(K\cdot F)_n)_- = B_{\rm min}}} \sum_{l=1}^L F_l^2
\end{split}
\end{equation}

\begin{figure*}[!ht]
\centering
\includegraphics[width=0.49\textwidth]{\figdir/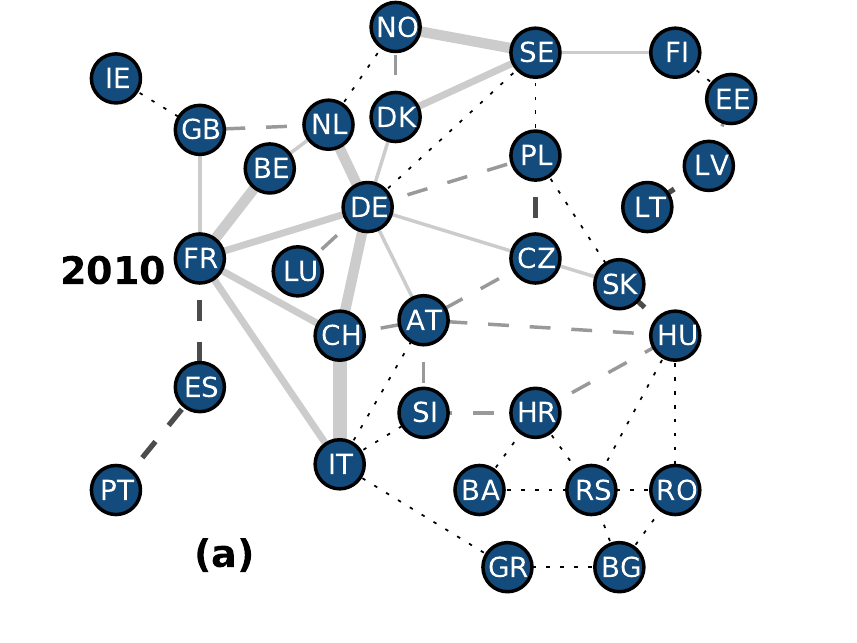}
\includegraphics[width=0.49\textwidth]{\figdir/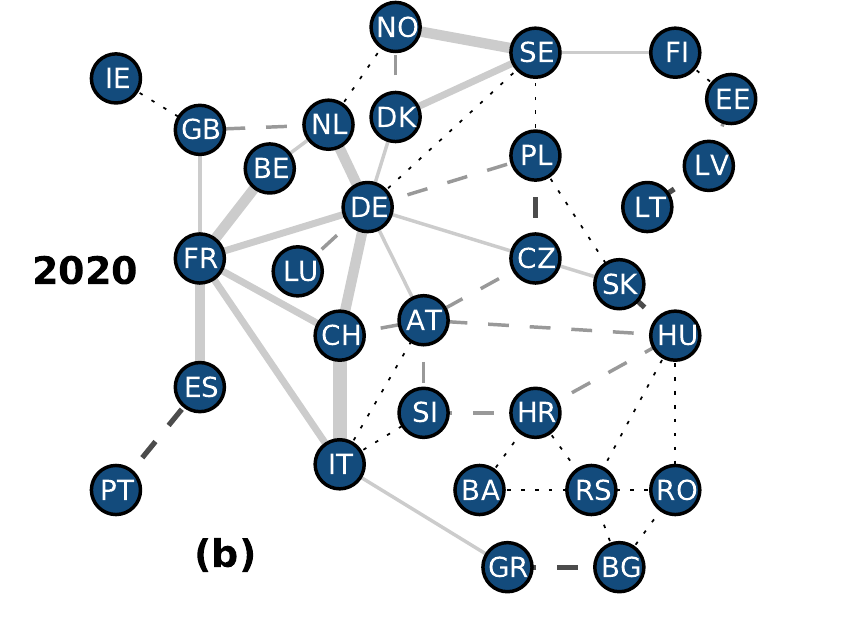}
\includegraphics[width=0.49\textwidth]{\figdir/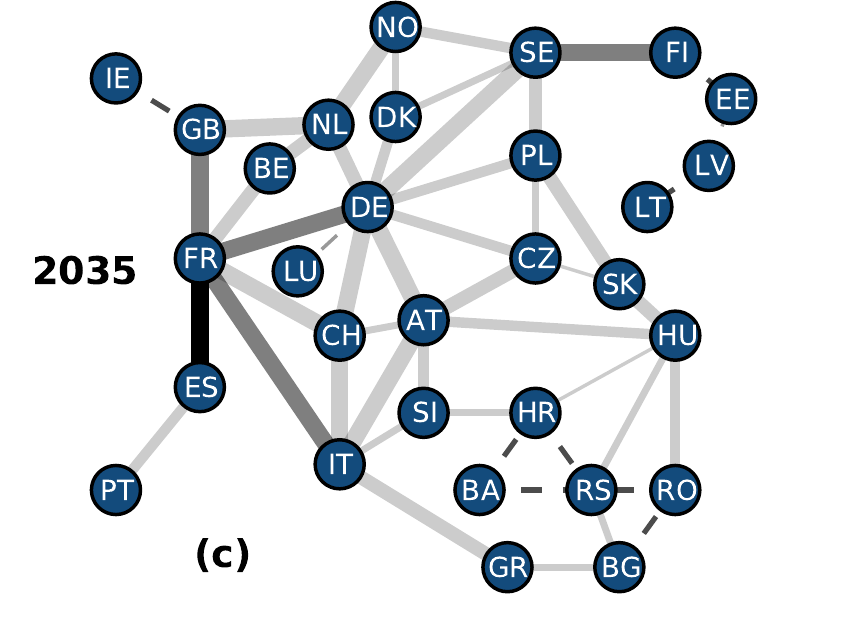}
\includegraphics[width=0.49\textwidth]{\figdir/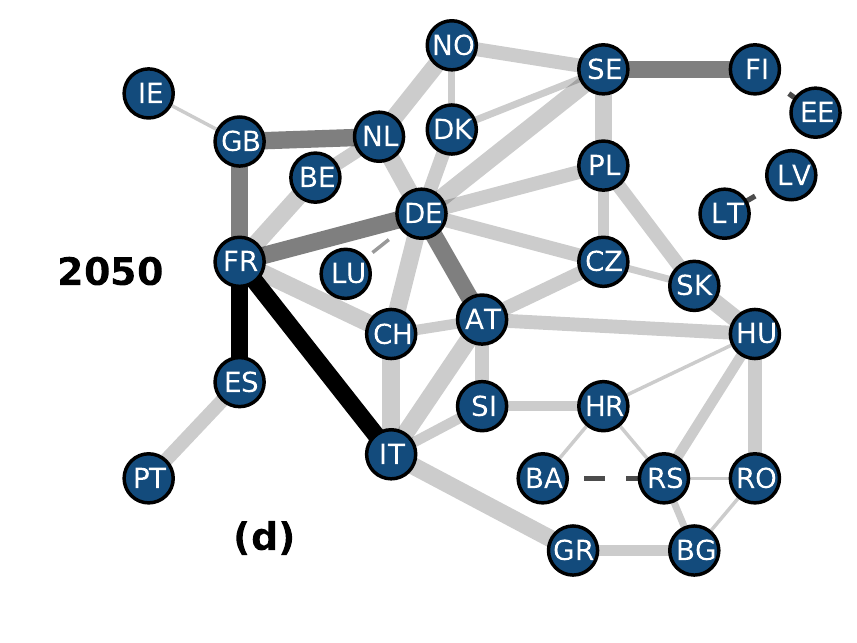}
\includegraphics[width=0.49\textwidth]{\figdir/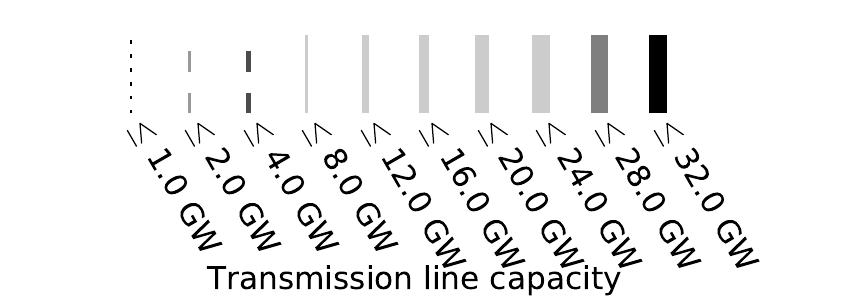}
\caption{
Network topology as used in our model: (a) present layout (winter 2010/2011) as
reported by ENTSO-E plus the three new links GB-NL, NO-NL, and SE-PL
\cite{entsoe_ntc,BritNed,NorNed,SwePol}, (b)-(d) development of the 90\,\% benefit of transmission
line capacities (see Sec.\ \ref{sec:trns7090}) for the years 2020, 2035, and
2050. Line style and thickness represents the larger one of the two net
transfer capacities of each link. Line lengths and node sizes are not to scale.
}
\label{fig:ntw}
\end{figure*}

The network topology employed is shown in Fig.\ \ref{fig:ntw}. We use five
different transmission layouts, i.e.\ line capacity constraints: Zero
transmission, transmission capacities as of today (shown in Fig.\
\ref{fig:ntw}a), unconstrained transmission \citep{rolando}, and two layouts
that are introduced and described in Sec.\ \ref{sec:trns7090}. See also Tab.\
\ref{tab:linecap}.

\section{Results}

\subsection{Time dependence of balancing energy for two fixed transmission layouts}
\label{sec:trnsfixed}

We first focus on the base scenario, i.e.\ the single country balancing optimal
mix. The impact of different end-point mixes will be discussed in Sec.\
\ref{sec:optmix}. The first two transmission layouts we investigate are constant
in time:
\begin{enumerate}
\item Zero transmission.
\item Today's transmission layout: Net transfer capacities as of
 winter 2010/2011 from ENTSO-E, plus the three new links BritNed, NorNed, and
 SwePol \citep{entsoe_ntc,BritNed,NorNed,SwePol}. 
\end{enumerate}

\begin{figure*}[!ht]
\begin{center}
\includegraphics[width=0.49\textwidth]{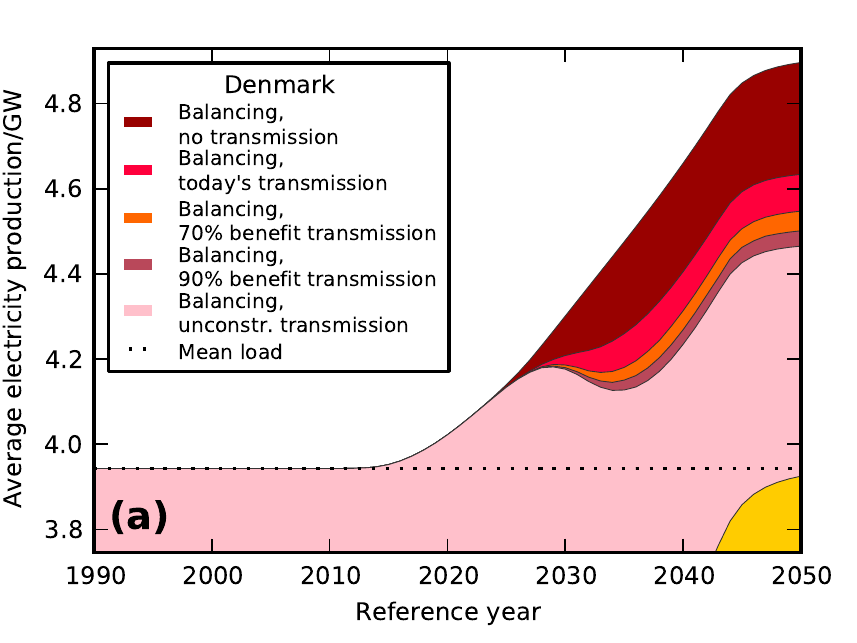}
\includegraphics[width=0.49\textwidth]{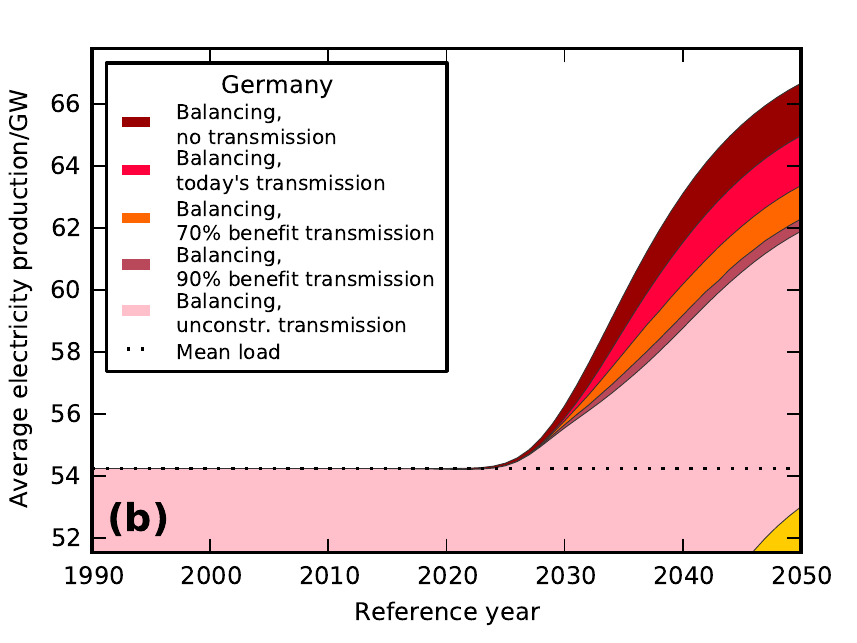}
\includegraphics[width=0.49\textwidth]{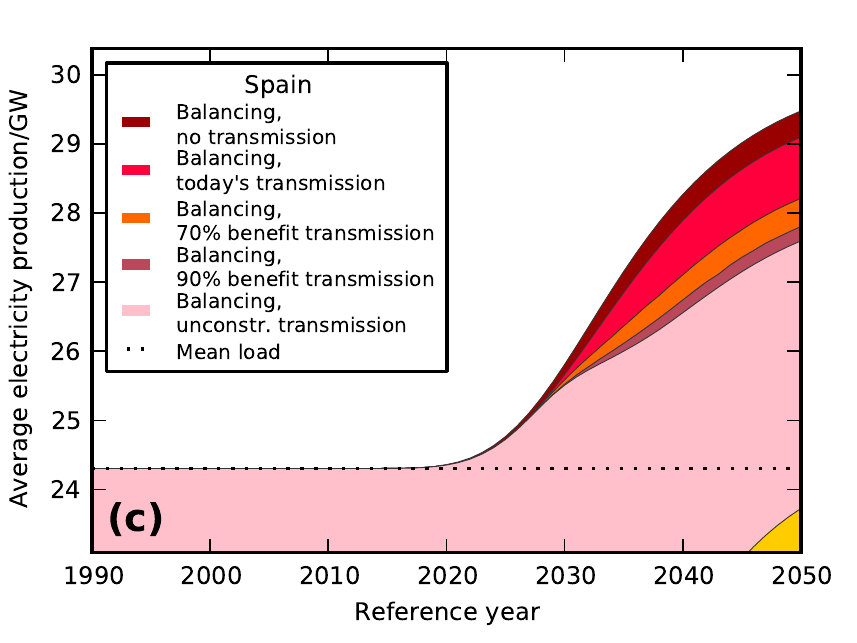}
\includegraphics[width=0.49\textwidth]{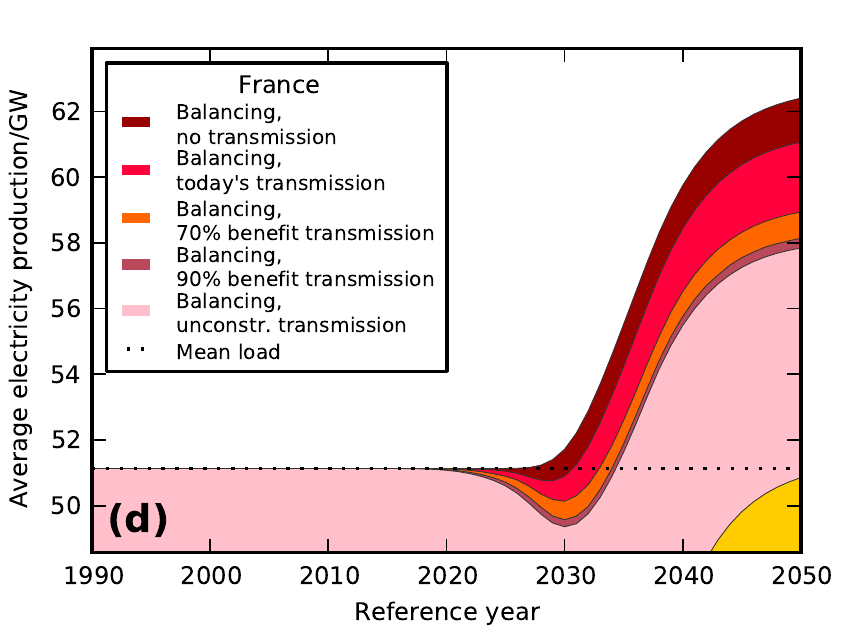}
\caption{
Zoom of Fig.\ \ref{fig:logfits}, now also showing the reduction of annual balancing
energy for various transmission layouts. For details on the transmission
layouts, see Secs.\ \ref{sec:trnsfixed}--\ref{sec:trns7090}. 
}
\label{fig:logfitszoom}
\end{center}
\end{figure*}

\begin{figure*}[!ht]
\centering
\includegraphics[width=0.490\textwidth,type=pdf,ext=.pdf,read=.pdf]{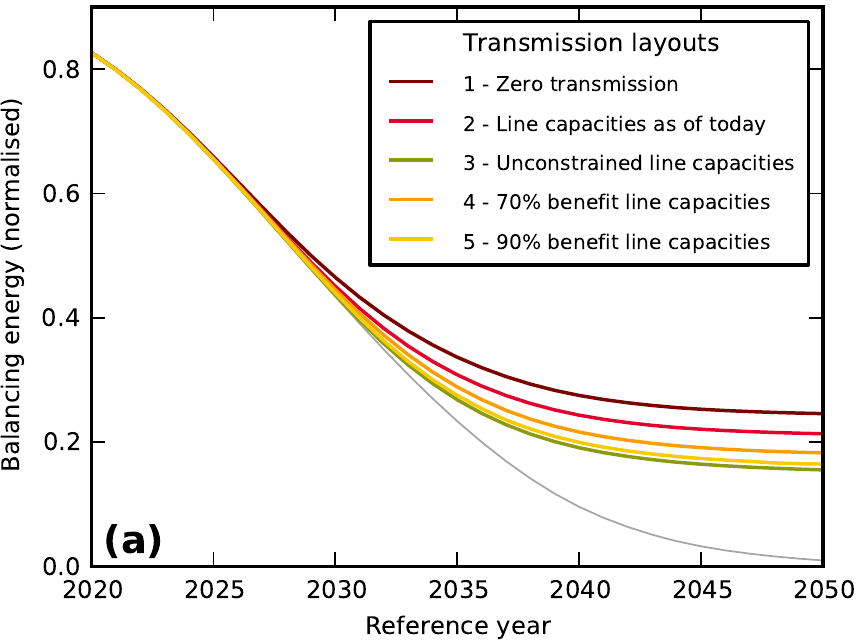}
\includegraphics[width=0.490\textwidth]{\figdir/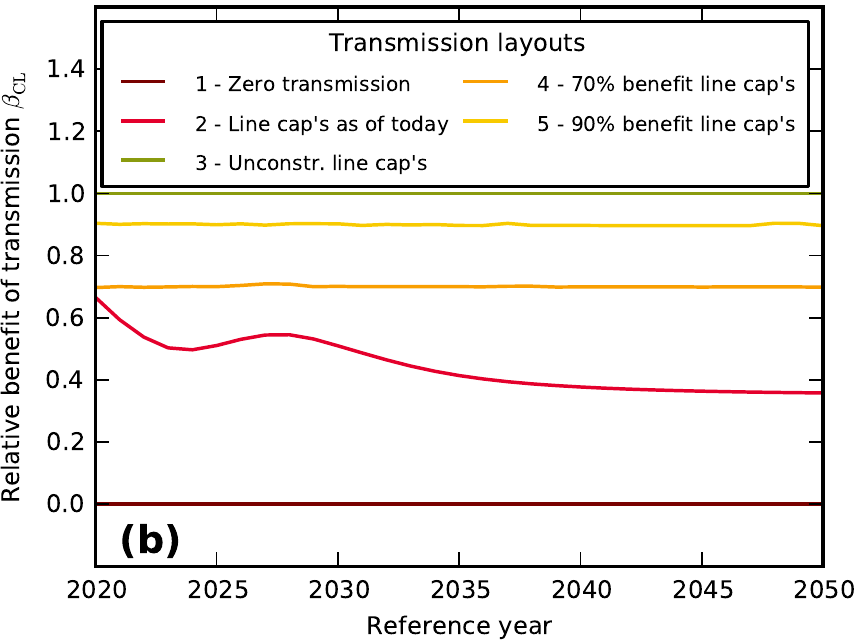}
\caption{
(a): Balancing energy vs. reference year for the various transmission layouts in
the base scenario. Note how the balancing is reduced by 70\,\% (layout 4) or
90\,\% (layout 5) of what is achievable by suitable transmission grid extensions
(see Sec.\ \ref{sec:trns7090}). This is also visible in panel (b), where the
relative benefit of transmission is shown, cf.\ Eq.\ \eqref{eq:benefit}. In panel
(a), the thin grey line shows the theoretical minimum balancing energy in each
reference year, i.e.\ the average normalised load minus the average
VRES penetration $\gamma$. Balancing energy is normalised by the mean load.
}
\label{fig:balred}
\end{figure*}

The resulting annual balancing energies for the single countries Denmark,
Germany, Spain, and France are shown in Fig.\ \ref{fig:logfitszoom}a-d, in dark
red for layout 1 and in red for layout 2. For each reference year, the annual
balancing energy has been averaged over the available 8 years of weather and
load data.  Balancing needs rise quite steeply for the case of no power
transmission, until it amounts to about 25\,\% of the average load in 2050. This
is mitigated slightly if line capacities as of today are assumed. As discussed
in more detail in Sec.\ \ref{sec:impexp}, single countries show different
intermediate behavior due to different trade opportunities.  The corresponding
figure for all of Europe is Fig.\ \ref{fig:balred}a, where balancing for layouts
1 and 2 is shown also in dark red and red, respectively.  

For comparison, Fig.\ \ref{fig:balred}a also shows the theoretical
minimum value for balancing as a thin, grey line. This would be obtained if the
entire VRES production could be used to cover the load, i.e.\ if no excess
production occurred. In this case, only a fraction of $(1-\gamma_\text{avg.}(t))
= (1 - \sum_n \gamma_n(t) \langle{L_n}\rangle / \langle{L_{EU}}\rangle)$ of the
total load needs to be covered from balancing. It can be seen that the
balancing needs of layout 1 depart already before 2030 from this optimal line,
while layout 2 follows the optimum up to 2030.

\subsection{Maximum reduction of balancing energy for the time-dependent unconstrained transmission layout}
\label{sec:trnsunconstrained}

\begin{figure}[!ht]
\centering
\includegraphics[width=0.49\textwidth]{\figdir/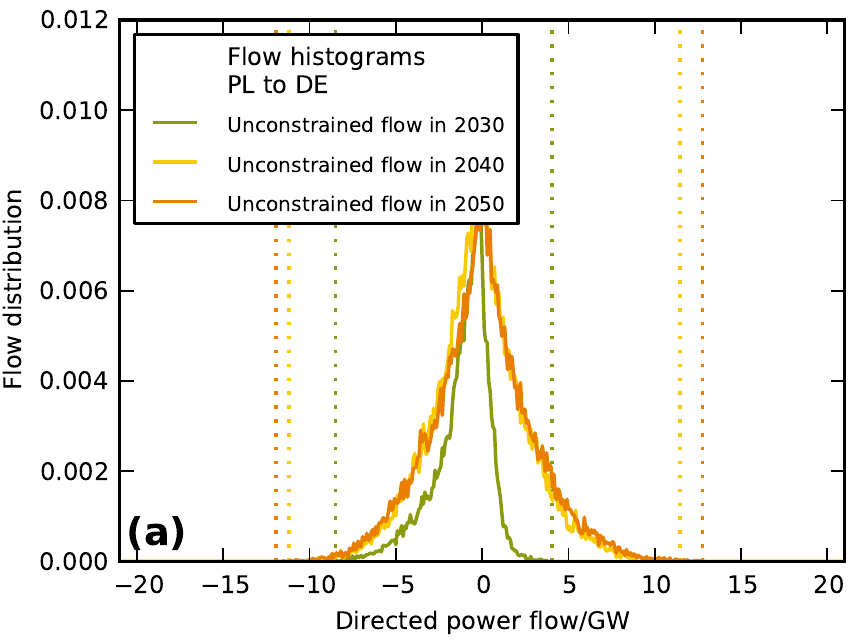}
\includegraphics[width=0.49\textwidth]{\figdir/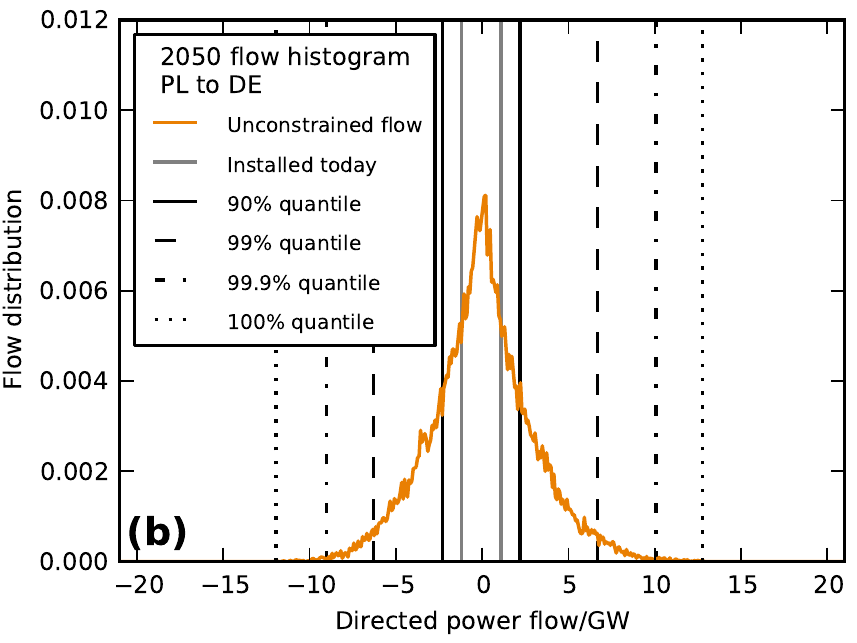}
\caption{
(a): Distributions of the unconstrained flow across the link between Germany and
Poland for the years 2030, 2040 and 2050. Only non-zero flows are shown. The
corresponding unconstrained line capacity layouts are assigned by setting the
capacity of a given link to the maximum value that occurs in the unconstrained
flow during the eight year data time series (dotted vertical lines). These
values can be found in the last four columns of Tab.\ \ref{tab:linecap}. 
(b): 2050 unconstrained flow for the same link, with (direction specific) quantiles. 
}
\label{fig:flwdistr}
\end{figure}

The next transmission layout is chosen to be:
\begin{enumerate}
\setcounter{enumi}{2}
\item Unconstrained transmission.
\end{enumerate}
It is possible to a posteriori associate a finite line capacity layout to
unconstrained transmission, simply by setting the link capacities to the maximum
value of the flow that is observed during the eight years of data, see 
Fig.\ \ref{fig:flwdistr}a. Detailed numerical values can be found in Tab.\
\ref{tab:linecap}. Since in this layout a single hour's flow determines the
capacity of a link, it sometimes happens that these capacities drop from one
reference year to the next for single links. This would correspond to a
downgrade of an already built link, which is unrealistic. Such artifacts have
therefore been removed, making the single links' capacities monotonously
increasing in reference year by keeping them at least at the levels reached in
previous years.

To see the effect of this layout on balancing energy, we look again at Figs.\
\ref{fig:balred}a (green line) and \ref{fig:logfitszoom}a-d (pink area). We see
that in Fig.\ \ref{fig:balred}a, balancing energy follows the theoretical
minimum (grey line) about five years longer than layout 1, up to about 2032.
Additionally, it is able to reduce the final balancing needs considerably, by
about 40\,\% of its value at zero transmission. It cannot, however, reduce
balancing energy down to zero: Power transmission is only able to match
surpluses at some nodes with deficits at others. If there is a global deficit
across all of Europe, balancing energy is needed no matter how strong the
transmission grid. In Fig.\ \ref{fig:logfitszoom}d, it is seen that balancing
for France drops temporarily such that combined own VRES generation and own
balancing become lower than the average load. This is due to imports of VRES and
will be discussed in more detail in Sec.\ \ref{sec:impexp}.

\begin{figure}[!ht]
\centering
\includegraphics[width=0.49\textwidth]{\figdir/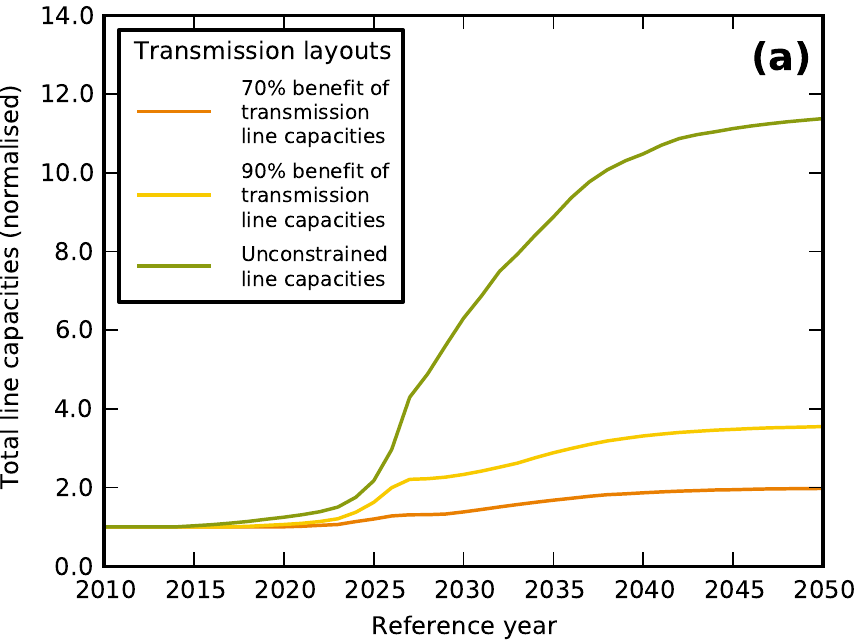}
\includegraphics[width=0.49\textwidth]{\figdir/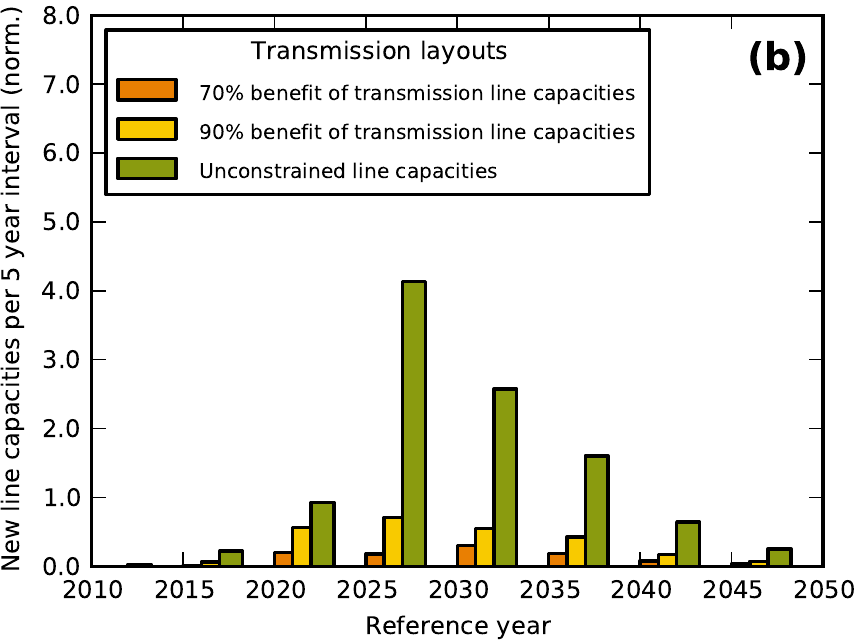}
\caption{
(a): Growth of total installed line capacity for the three time-dependent line
capacity layouts unconstrained transmission (layout 3), 70\,\% benefit of
transmission (layout 4), and 90\,\% benefit of transmission (layout 5). It is
seen that layout 3 requires a final installation of almost twelve times of what
we have today, layout 4 about twice as much, and layout 5 a little less than
four times as much. 
(b): Five-year increment of line installations for the same
three layouts. In both panels, the total installation is normalised by the
installation of today (see Sec.\ \ref{sec:trnsunconstrained} for details).
}
\label{fig:transinvest}
\end{figure}

Fig.\ \ref{fig:transinvest}a,b show in green the total necessary line upgrades to
obtain layout 3. Fig.\ \ref{fig:transinvest}a depicts the total line
capacities that need to be installed, and Fig.\ \ref{fig:transinvest}b
shows the increments per five-year interval. For calculating the total
transmission capacity, the larger one of the two NTC values of each link is used
as a proxy to its physical capacity. These yield a sum of approximately $74\,\rm
GW$ for the total line capacities installed today.  The necessary total line
capacities are plotted as multiples of this number. It is seen that line
capacities for this layout would amount to almost twelve times of today's
installation in the end, and require a top installation speed of roughly adding
today's installation each year between 2025 and 2030.

\subsection{Two compromise transmission layouts}
\label{sec:trns7090}

While the absence of new line investment of the two fixed layouts 1
and 2 makes them attractive, they lead to large balancing needs.
Balancing needs are considerably reduced by the unimpeded power exchange of the
unconstrained layout 3. This, however, requires huge investments in reinforced
transmission lines. The idea is now to choose a line capacity layout which yields
a certain reduction of balancing needs while keeping new line investments in a
reasonable range.

The benefit of transmission for balancing reduction can be quantified in the
following way: Denote a capacity layout by $C_\text{L}$, representing
a set of net transfer line capacities. The total balancing $B_\text{tot}(C_\text{L})$
can then be understood as a function of $C_\text{L}$. The two
extreme cases are the zero capacity layout $C_\text{zero}$ (layout 1), where all
transmission capacities are set to zero, and the unconstrained layout
$C_\text{unconstrained}$ (layout 3), where the transmission capacities are
determined from what is necessary for unimpeded flow. The relative benefit of
transmission $\beta(C_\text{L})$ of a generic layout $C_\text{L}$ can then be 
expressed as the reduction of balancing achieved by installing $C_\text{L}$ 
divided by the maximum possible benefit of transmission, which is obtained when 
switching from $C_\text{zero}$ to $C_\text{unconstrained}$: 
\begin{align}
  \beta(C_\text{L}) 
    =  \frac{B_\text{tot}(C_\text{zero})-B_\text{tot}(C_\text{L})}
               {B_\text{tot}(C_\text{zero})-B_\text{tot}(C_\text{unconstrained})}
        \; .
\label{eq:benefit}
\end{align} 
\
By construction, the relative benefit of transmission is zero for the zero capacity layout 1, 
and it is one for the unconstrained capacity layout 3. Fig.\ \ref{fig:balred}b illustrates 
the relative benefit of transmission for today's capacity layout 2. It decreases with
progressing reference years and converges from above to about $\beta(C_\text{today})=0.34$ 
for the final reference years. 

Two new compromise transmission layouts are:
\begin{enumerate}
\setcounter{enumi}{3}	
\item 70\,\% benefit of transmission capacities: For all links, fix a quantile of
  the unconstrained 2050 flows as transmission capacity (cf.\ Fig.
  \ref{fig:flwdistr} (b)), such that 70\,\% of the relative benefit of transmission
  is harvested (cf.\ Fig.\ \ref{fig:balred} (a) and (b)).
\item 90\,\% benefit of transmission capacities: Analogously to 70\,\%
  benefit of transmission capacities.
\end{enumerate}
There are a multitude of possible interpolations between zero transmission and
unconstrained transmission that lead to the same desired balancing energy
reduction and benefit of transmission as depicted in Fig.\ \ref{fig:balred}a+b.
We choose the quantiles of the corresponding end-point (2050) unconstrained flow
distribution because this policy has been found to perform best in terms of
minimal transmission installations and maximum balancing reduction in
\cite{rolando}. For an illustration of the flow quantiles, see Fig.\
\ref{fig:flwdistr}b for the link between Poland and Germany. We work with
quantiles of the end-point flows rather than with quantiles of the unconstrained
flow of the same year in order to consistently build up the capacity layout that
is actually needed in the end. The quantile line capacities are capped by the
unconstrained capacities of the corresponding years in order to avoid a
premature installation of lines that are not used immediately. Furthermore,
layouts 4 and 5 start from today's NTCs, that is, no dismantling of existing
lines is assumed.

The total transmission line capacities required to obtain the 70\% and 90\%
benefit of transmission and the amount of new installations per five year
intervals are shown in Fig.\ \ref{fig:transinvest}a and b. In addition, the
capacity of each single link in the years 2020, 2030, 2040, and 2050 in the base
scenario for the year-dependent line capacity layouts 3, 4, and 5 can be found
in Tab.\ \ref{tab:linecap}. The build-up of the 90\,\% benefit of transmission
layout is also illustrated in Fig.\ \ref{fig:ntw}. About two and four times as
much as what is installed today is needed to harvest 70\,\% and 90\,\% of the
possible benefit of transmission, respectively. Both schemes seem within reach.
For single links, the 90\,\% benefit layout agrees nicely with the results of
\cite{energynautics_grids}, which finds e.g.\ a line capacity of $19.5\,\rm GW$
between Spain and France for the fully renewable stage, compared to $17.3\,\rm
GW$ found in our base scenario.  For the link between France and the UK, they
report a final capacity $6\,\rm GW$, while we find $11.9\,\rm GW$. But this is
due to the different grid topology they use, which includes an additional link
from Great Britain to Norway with a capacity of $3.5\,\rm GW$, and a substantial
offshore grid in the North Sea with a total capacity of $16.5\,\rm GW$.

Notably, the build-up in line capacities has to start no later than 2020 if the
desired 70\,\% or 90\,\% benefit of transmission is to be harvested throughout
the years. We have to keep in mind that the absolute reduction in balancing is
small in the beginning (cf.\ Fig.\ \ref{fig:balred}), such that the total losses
from not building the lines would be small at first. But, as balancing needs
grow, the need for transmission capacity quickly increases. So it would be 
advisable to start the line build-up as soon as possible.

We take a combined look at Tab.\ \ref{tab:logfit} and Figs.\ \ref{fig:logfits},
\ref{fig:balred} and \ref{fig:transinvest}. Fig.\ \ref{fig:logfits} and Tab.
\ref{tab:logfit} show that the main part of the wind installation growth takes
place 2015-2035, while solar PV installations are a little later, about
2020-2040. Figs.\ \ref{fig:logfits} and \ref{fig:balred}a show an interesting 
feature: The onset of additional balancing energy beyond the minimum of 
$(1-\gamma_\text{avg.})$ times the average load can be postponed by transmission
by five years.  In order to achieve this, the main transmission line growth has
to happen from 2025 to 2030, leading to the peak in new installations seen in
Fig.\ \ref{fig:transinvest}b at that time.

\subsection{Import and export opportunities}
\label{sec:impexp}

It is possible to investigate the effects of a strong
transmission grid on the import and export opportunities of single countries. We
only consider trade with VRES since other forms of electricity generation are
not treated explicitly in our model. The transition at different times in
different countries has some interesting effects. There are roughly three
coordinates which determine export and import opportunities of a country, namely
the size of its mean load, its position in the network (central or peripheral)
and the time of transition to VRES with respect to its neighborhood. 
\begin{figure*}[!ht]
  \centering
    \includegraphics[width=0.49\textwidth]{\figdir/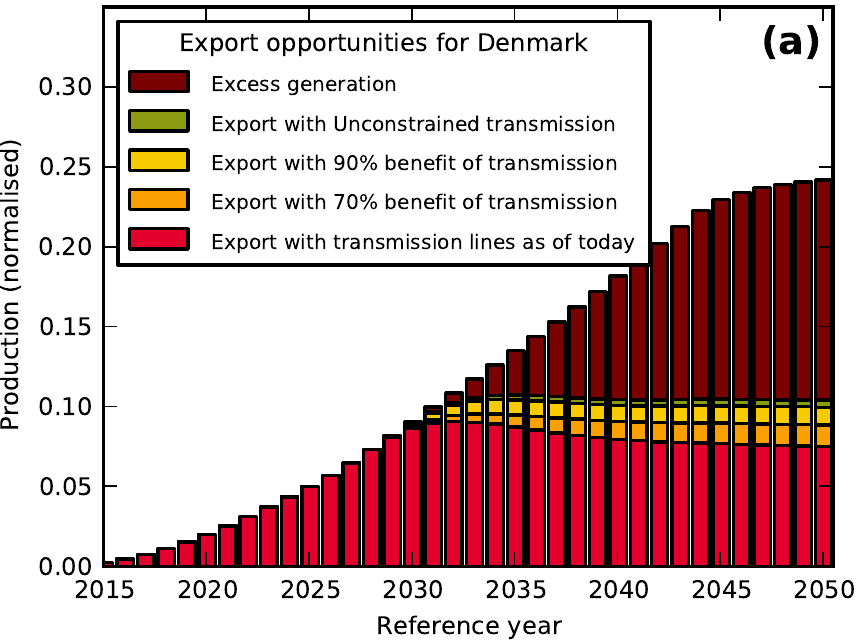}
    \includegraphics[width=0.49\textwidth]{\figdir/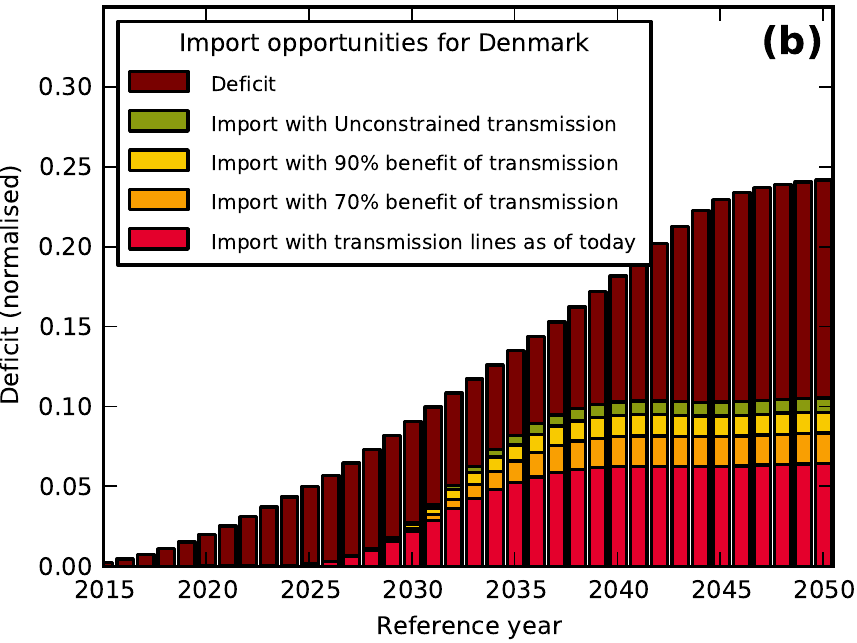}
    \includegraphics[width=0.49\textwidth]{\figdir/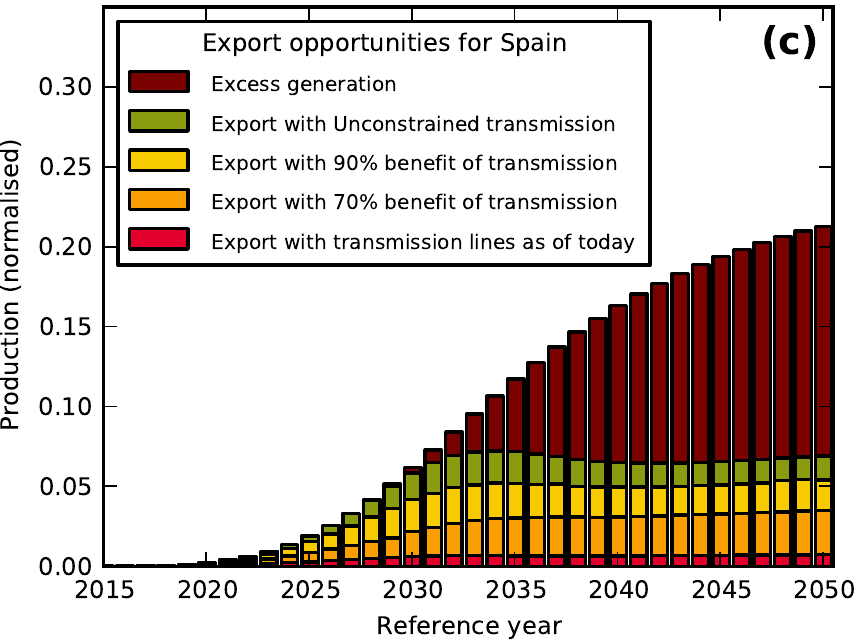}
    \includegraphics[width=0.49\textwidth]{\figdir/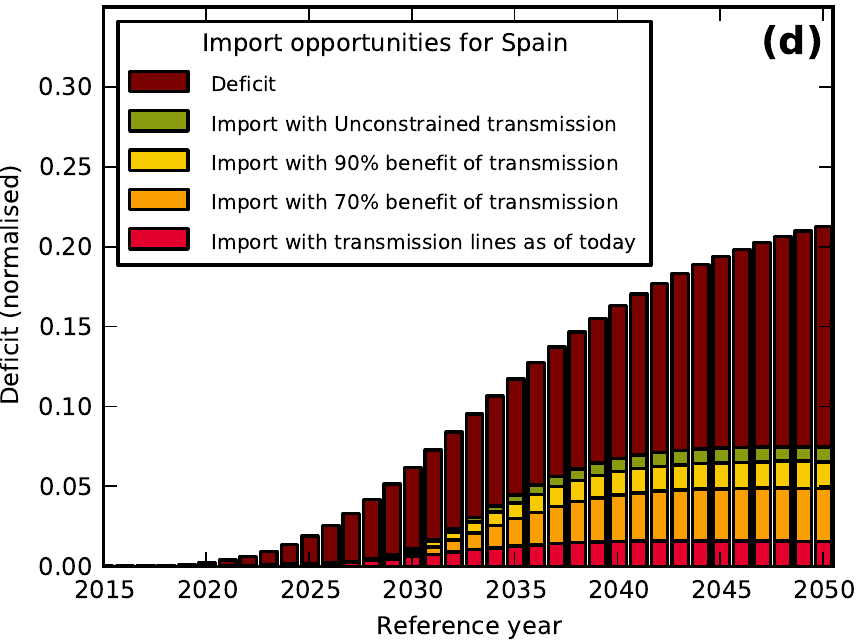}
    \includegraphics[width=0.49\textwidth]{\figdir/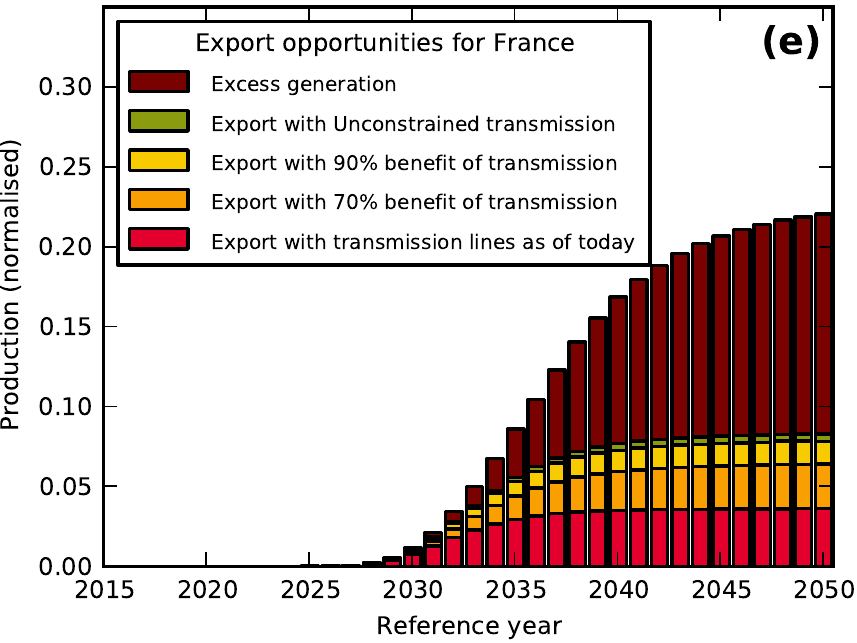}
    \includegraphics[width=0.49\textwidth]{\figdir/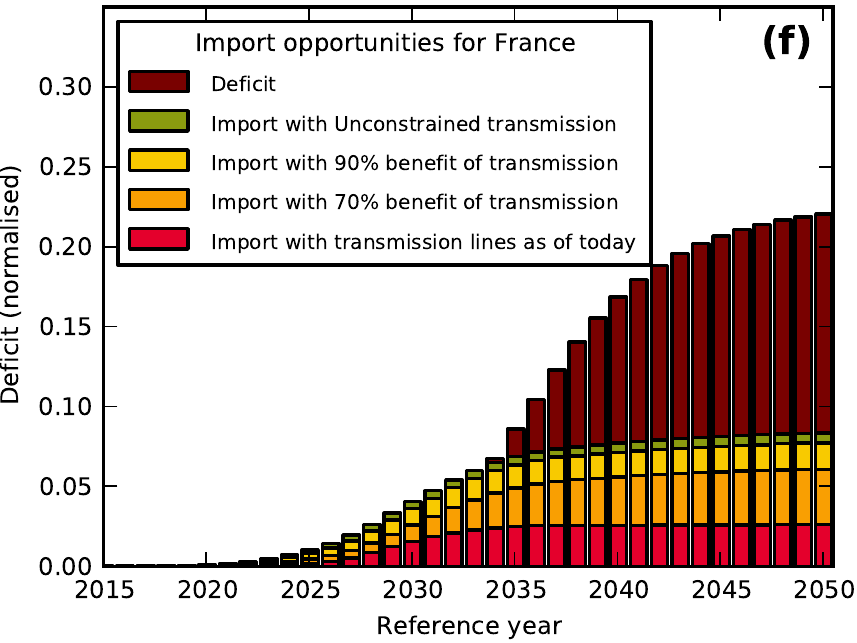}
  \caption{Export and import opportunities for selected countries, in the base
scenario. Panels (a), (c), and (e) show the export opportunities in case of a
domestic overproduction for Denmark, Spain, and France, respectively. Panels
(b), (d), and (f) depict the import opportunities for the same countries, which
are used to cover a domestic deficit or to replace domestic balancing.}
\label{fig:expimp} 
\end{figure*} 

\subsubsection*{Load size} A country's load size influences the ratio of imported or
exported energy to its deficit or surplus. A country with a high absolute load
will experience high deficit as well as high excess. If its neighbours have a
substantially smaller load, they will in most cases neither be able to absorb a
large excess nor cover a large deficit. This means that the fraction of the
deficit that can be covered by imports and the fraction of the excess that can
be exported is smaller the larger a country's load.

\subsubsection*{Position} The position of a country in the network is another factor that
influences its trade opportunities. This point is a little more subtle and
partly due to our power flow modelling which minimises the overall flow in the
network. Neglect for a moment restricted transmission line capacities. Imagine a
situation where some countries are in deficit and some are in excess, such that
there is an overall shortage. The question is now which of the countries with a
deficit gets the excess from those which see a surplus production. In a specific
situation, the answer depends on the details of the distribution of excess and
deficit in the network, but on average, transport of energy to a remote country
causes more flow than to a central country, such that the central country will
be preferred. The same goes for a situation where there is global excess, but
some countries with a deficit: Central countries have a higher chance of
exporting than peripheral ones, because this will on average cause less flow.
Taking now limited transmission capacities into account, the situation is
accentuated: A lot of flow to or from peripheral countries is not only
suppressed by the flow minimisation, but may be altogether impossible.

\subsubsection*{Time of transition} Whether the transition to VRES occurs early or late in
a country does not have an effect on the end-point import/export capabilities,
but becomes important during the transition. If the transition in a country
takes place early, it experiences deficit and excess situations earlier than its
neighbours. For the first years, this means that in case of a deficit,
neighbours are probably not able to export anything because they do not see
excesses yet. On the other hand, because VRES generated electricity is shared
wherever possible, if the early country has a surplus production, it can almost
certainly export it to later neighbors, where it replaces local balancing.
In short, an early transition may mean poor import opportunities, but on the
other hand good export opportunities during the first years. These differences
are subsequently diminished as all countries switch to VRES-based electricity
supply, and then size and position become the dominant factors determining
import and export opportunities.

In the reference year 2050, all countries reach a VRES share of (almost) $\gamma
= 1$. As shown in \cite{rolando}, even unlimited transmission can reduce the
total balancing in Europe by only 40\,\% in the $\gamma=1$ case, due to the
spatio-temporal correlations in the weather. This implies that in 2050, only
40\,\% of the total deficit can be covered by imports and equally only 40\,\% of
the total excess can be exported. This statement is valid for the load-weighted
average over all countries. Single countries see deviations due to their
position and load size as explained above.

We take a look at three examples from the different classes which arise from
these distinctions. We start with Denmark, which is small, central and an early
adopter; see Figs.\ \ref{fig:expimp}a and b as well as Fig.\
\ref{fig:logfitszoom}a. Denmark's wind power installation covers on average more
than 33\,\% of the load already today \cite{danmark_wind}. Up to now, the excess
production can easily be exported into the neighbouring countries. On the import
side, at first there are no neighbours willing to export any VRES generation
because they can use everything domestically.  This changes quickly as soon as
the neighbours catch up with their VRES installation causing them excess
production. Since the neighbours have a larger total production, resulting in
more excess, the import opportunities are actually very good then, even leading
to a dip in balancing energy between 2030 and 2040 for reinforced transmission
grids, see Fig.\ \ref{fig:logfitszoom}a. Between 2045 and 2050, finally all
countries reach a VRES share close to 100\,\%. This means that export
opportunities are reduced: Since import can only replace balancing, but not
domestic VRES production, it becomes less probable to find a customer for excess
production.

For comparison, we now look at Spain (Figs.\ \ref{fig:expimp}c and d and Fig.\
\ref{fig:logfitszoom}c), which also has ambitious VRES targets for the near future,
but is peripheral. There is only one strong connection to Portugal. This
connection does not improve the import/export capabilities of Spain much,
since Portugal's load is less than one fifth of the Spanish load, and
it can therefore not absorb much of the Spanish fluctuations. We see that Spain
starts similarly to Denmark with good potential export opportunities and no good
chances to cover its deficits by import. However, there is a significant
difference between the transmission capacities needed: While for Denmark today's
transmission capacities are already sufficient to export most of its excess, the
weak link from Spain to France blocks almost all export if it is not reinforced.
The same is true to a lesser degree for the import opportunities. In the further
development, the import evolves appreciably different: Compared to Denmark,
Spain's import opportunities are poor. This is due to two reasons: Firstly,
Spain has a bigger mean load. Its total deficit is therefore larger and harder
to cover. Secondly, it is peripheral. In cases where there is only an
insufficient supply of excess power which some countries want to export, while
Spain and other countries have a deficit, the export flow will probably dry out
before reaching Spain.

As a last example, we look at a central country which is large and relatively
late, namely France (Figs.\ \ref{fig:expimp}e and f and Fig.\
\ref{fig:logfitszoom}d). Its import opportunities are reasonably good for a
country with a large load, but could be significantly improved by increased
transmission capacities. The same holds for the export. As the transition to
VRES is expected to be rather late in France, there are no deficits that cannot
be covered by imports during the first years.  In fact, VRES imports can already
be used to replace balancing even before there is a significant domestic VRES
installation, causing France to produce less than its own load from VRES and
balancing. The rest can be covered by VRES imports (see the dip in Fig.\
\ref{fig:logfitszoom}d). On the other hand, France does not experience an export
boom in the beginning.

\subsection{Shift of optimal mix}
\label{sec:optmix}

\begin{figure}[!ht] \centering
\includegraphics[width=0.49\textwidth,type=pdf,ext=.pdf,read=.pdf]{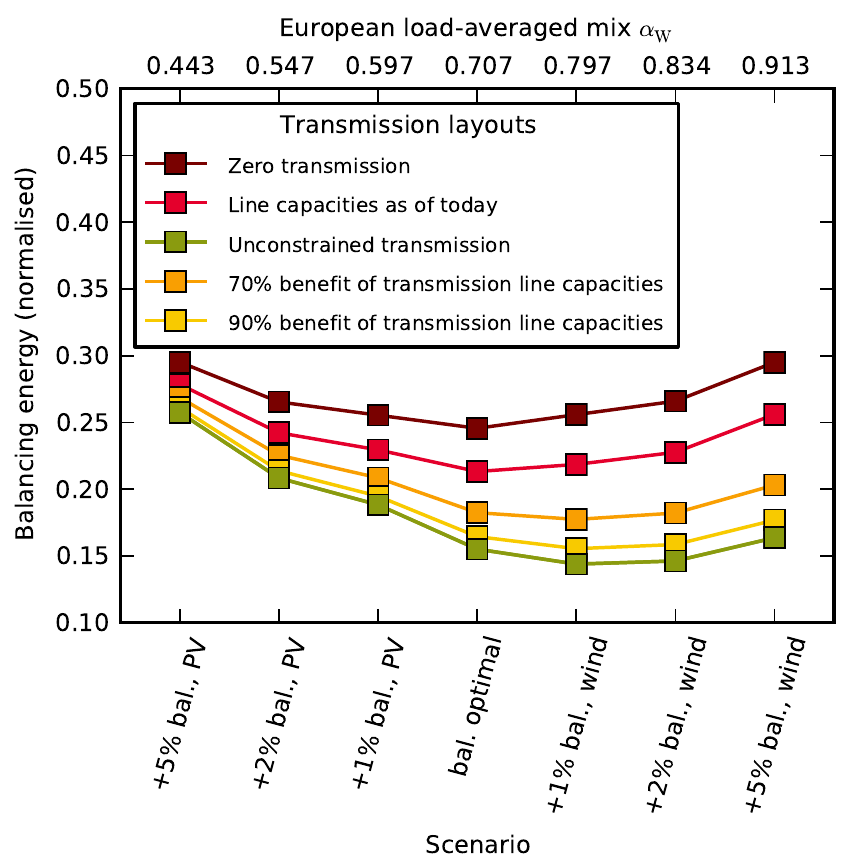}
\caption{
Annual European balancing energy for different average end-point mix 
scenarios for the year 2050. Note that the average mix can be read off 
the nonlinear top axis. Balancing is normalized by the average load.
}
\label{fig:mix_sweep}
\end{figure}

So far we have used only the base scenario for the logistic growth of wind and
solar power generation, where the 2050 end-point mix of each country is chosen
such that it minimises the average balancing energy based on the zero
transmission capacity layout. At the end of Sec.\ \ref{sec:logfit}, six
additional mix scenarios have been defined, with three of them having an
increasingly larger share of wind power generation and the other three having an
increasingly larger share of solar power generation. We will now use these six
additional scenarios and investigate the impact of different relative mixes
between wind and solar power generation on the combined balancing and
transmission needs.

\begin{figure*}[!ht]
\centering
\includegraphics[width=0.49\textwidth,type=pdf,ext=.pdf,read=.pdf]{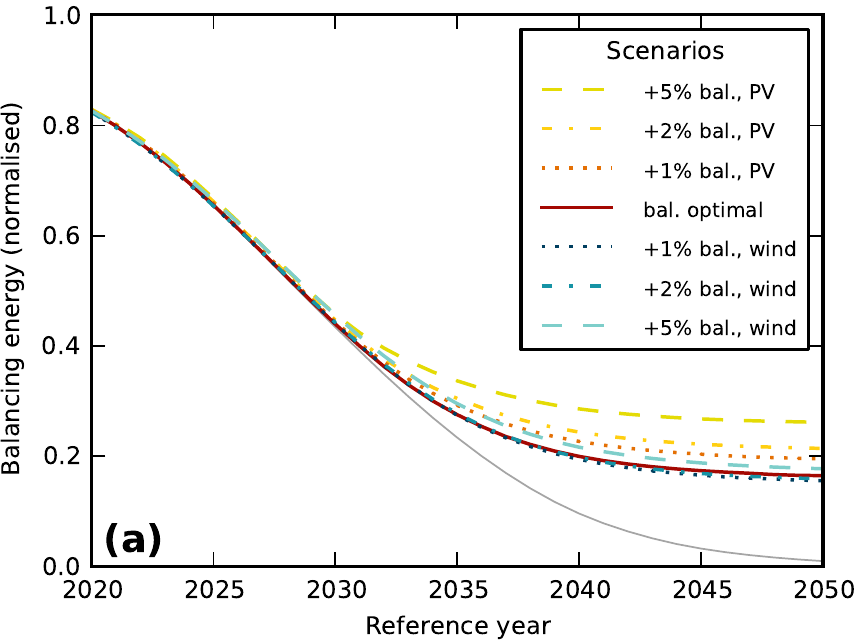}
\includegraphics[width=0.49\textwidth,type=pdf,ext=.pdf,read=.pdf]{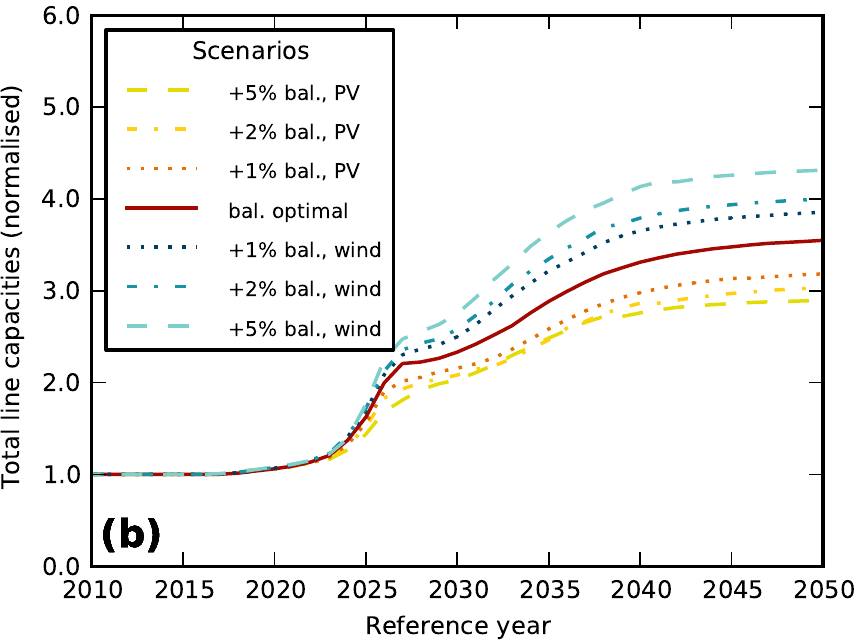}
\caption{
(a) Balancing energy and (b) total net transfer capacities vs.\ reference year
for the base scenario, the three wind heavy and the three solar heavy scenarios,
based on the 90\,\% benefit of transmission line capacities (layout 5). Total
net transfer capacities have been normalized by the total installation of today.
The dark red curves for the base scenario in panels (a) and (b) are identical to
the yellow curves in Figs.\ \ref{fig:balred}a and \ref{fig:transinvest}a,
respectively.  }
\label{fig:capinv}
\end{figure*}

Fig.\ \ref{fig:mix_sweep} shows the dependence of the annual European balancing
energy on the different mix scenarios and on the different transmission layouts
for the final reference year 2050. Once transmission is introduced, a higher
wind share performs better in terms of balancing reduction. Compared to the base
scenario, the two wind heavy $+1\,\%$ and $+2\,\%$ scenarios result in a lower
European-wide balancing energy once the strong transmission layout types 3-5 are
considered. The result is consistent with \cite{rolando}, where an optimal
end-point mix $\alpha^{\rm W}_\textup{agg.}=0.822$ was found for an aggregated
Europe with an unconstrained transmission capacity layout. The underlying reason
for the shift towards more wind when large regions are interconnected is that
the spatial correlation of wind power generation drops significantly over
distances of 500 to 1000 km, while solar PV remains more correlated. The effect
is well illustrated for the case of Sweden in \cite{Widen:2011ys}. When
comparing the total balancing energy required in the base scenario and the
$+1\,\%$ and $+2\,\%$ wind heavy scenarios, it becomes clear that the absolute
difference between the three is relatively small for the strong transmission
capacity layouts.  This is again demonstrated in Fig.\ \ref{fig:capinv}a, which
shows the balancing energy as a function of the reference year, based on the
90\,\% benefit of transmission line capacities (layout 5). 

Fig.\ \ref{fig:capinv}b illustrates the development of the total line capacities
and reveals that a high wind share leads to higher transmission needs. This is
due to the fact that for a high wind share, there is a high chance of covering a
shortage in one country with excess from another. Solar PV power output, on the
other hand, is much more correlated, being bound to the day-night pattern.
Therefore, in case of a high share of solar PV, it is less probable that deficit
and surplus production can cancel each other, implying less transmission needs. 

The impact of the different scenarios for the relative mixes between wind and
solar power generation on the combined balancing and transmission needs is
summarised in Tab.\ \ref{tab:devs}. An opposing trend is observed: whereas the
balancing energy decreases with an increasing, not-too-large share of wind power
generation, it is the opposite for the total transmission capacity needs. This
finding shows that when determining the complex features of a fully renewable
energy system for Europe, it is not sufficient to look at isolated regions only,
but also at the interplay between regions.

\begin{table*}[!t]
\caption{
Balancing energy needs and total transmission investment in 2050 resulting for 
different end-point mix scenarios, based on the 90\,\% benefit of transmission 
line capacities. All quantities are given as the relative deviation from the base 
scenario.
}
\label{tab:devs}
\centering
\begin{tabular}{lrr}
\hline
Scenario & Change in balancing energy & Change in transmission investment\\
\hline
$+5\,\%$ balancing, PV    & $58.8\,$\%  & $-18.9\,$\% \\
$+2\,\%$ balancing, PV    & $30.0\,$\%  & $-14.8\,$\% \\
$+1\,\%$ balancing, PV    & $18.5\,$\%  & $-10.6\,$\% \\
base                                 &   $0.0\,$\%  &    $0.0\,$\% \\
$+1\,\%$ balancing, wind &  $-5.3\,$\%  &    $9.1\,$\% \\
$+2\,\%$ balancing, wind &  $-3.5\,$\%  &  $13.1\,$\% \\
$+5\,\%$ balancing, wind &   $7.9\,$\%  &   $22.0\,$\% \\
\hline
\end{tabular}
\end{table*}

\section{Conclusions}

Using the logistic growth as an assumption for how the share of VRES in Europe
grows from today up to a fully VRES-supplied European power system, we have
investigated the required amount of transmission capacities that allow most of
the excess from VRES to be shared across Europe during the transition. Most of
the wind installation growth is expected to occur between 2015 and 2035, while
the solar PV growth period is shifted by about five years. As soon as the total
VRES share reaches levels of about 35\,\% in 2025, surplus production starts to
rise without a strong transmission grid. This can be postponed by up to five
years by transmission reinforcements, until the average VRES penetration has
grown to more than 50\,\%. This line build-up has to start strongly during the
early years from 2020 to 2030, and has to be continued on similar levels for
fifteen more years in order for transmission to have a large overall system
benefit. Such a transmission line investment is large, but appears to be
feasible since the overall installation needs to reach only about two to four
times of what we have today.

Reinforced transmission lines have several effects. First\-ly, without any
demand-side management and storage, transmission alone leads to a significant
reduction in balancing energy needs by about one third.  Secondly, the import
and export opportunities become better for all countries, especially for those
that are peripheral and go early for a high share of VRES, such as Ireland and
Spain. Thirdly, the optimal mix of wind and solar changes: With more
transmission, wind grid integration becomes easier, and so the optimum shifts
from about a 70\,\% wind share to about 80\,\% wind share.

Import and export of electricity is greatly facilitated by a reinforced
transmission grid. It has to be kept in mind, however, that while countries
which increase their VRES penetration early benefit from the strong transmission
lines during the first years, the export and import opportunities decrease once
all countries have reached a high VRES share. This problem requires more
attention in future research: most likely it can  be diminished by the use of
country-internal smart-energy systems, including  storage, demand-side
management, and conversion of excess electricity to heat.

This study is open for several possible future extensions. New transmission
lines can be added to the topology of the employed network, which are either
being built right now or planned in the near future, e.g.\ between Norway and
Great Britain. New nodes can be added, reflecting Europe's connections its
neighbours. Dispatchable renewable sources, such as biomass and hydropower,
could be included explicitly into the system. Note however, that the inclusion
of hydro power in particular will lead to more transmission requirements, since
the sites at which it is available are mainly Norway, Switzerland, and Austria.
Taking these ideas further, a multi-dimensional optimization of the electricity
supply can be investigated, including heterogeneous line capacities, balancing
power capacities, balancing energy reserves, and renewable generation mix.
Storage would be included as well, splitting it up into a local, small-capacity
part, representing battery storage or demand-side management, and a large
storage of seasonal capacity, representing e.g.\ hydrogen storage in salt
caverns in Northern Germany, similar to the approach of \cite{morten}. This
localized large storage facility will again lead to more power flow.

\section{Acknowledgements}

SB gratefully acknowledges financial support from O. and H. St\"ocker, and GBA
from DONG Energy and the Danish Advanced Technology Foundation. Furthermore, we
would like to thank Uffe V. Poulsen and Tue V. Jensen for helpful discussions.

\section*{References}
\bibliographystyle{unsrt}
\bibliography{literatur}

\onecolumn
\begin{appendix}
\label{sec:app}

\begin{small}

\section{Data tables}

\begin{center}
\begin{longtable}{l||r|ccc|c|ccc}
\caption{
End-point wind/solar mixes for three wind heavy, three solar heavy, and the
single country optimal mix base scenario, expressed as the relative share
$\alpha^{\rm W}$ of wind in VRES. The wind resp. solar heavy scenarios lead to
balancing needs increased by 1\,\%, 2\,\% and 5\,\% of the average load when
transmission is not included. The second last row shows the load-weighted
average of the mixes over all single countries, and the last row shows the
corresponding mixes for an aggregated Europe, where load and generation are
shared without transmission limits.
}\\
\label{tab:endmix}
& &  \multicolumn{7}{c}{\bf wind fraction $\alpha_{\rm W}$} \\
{\bf country} & {\bf avg. load/} &\multicolumn{3}{c|}{\bf solar heavy scenarios} & {\bf base}
& \multicolumn{3}{c}{\bf wind heavy scenarios} \\
& {\bf GW} & $+5$\,\% bal. & $+2$\,\% bal. & $+1$\,\%
bal. & opt. mix & $+1$\,\% bal. & $+2$\,\%
bal. & $+5$\,\% bal. \\
\hline
\endfirsthead
& &  \multicolumn{7}{c}{\bf wind fraction $\alpha_{\rm W}$} \\
{\bf country} & {\bf avg. load/} &\multicolumn{3}{c|}{\bf solar heavy scenarios} & {\bf base}
& \multicolumn{3}{c}{\bf wind heavy scenarios} \\
& {\bf GW} & $+5$\,\% bal. & $+2$\,\% bal. & $+1$\,\%
bal. & opt. mix & $+1$\,\% bal. & $+2$\,\%
bal. & $+5$\,\% bal. \\
\hline
\endhead
AT & 5.8 & 0.405 & 0.511 & 0.562 & 0.674 & 0.764 & 0.799 & 0.874 \\
\rowcolor[gray]{0.8} BA & 3.1 & 0.413 & 0.523 & 0.574 & 0.683 & 0.765 & 0.797 & 0.872 \\
BE & 9.5 & 0.426 & 0.523 & 0.577 & 0.701 & 0.802 & 0.837 & 0.911 \\
\rowcolor[gray]{0.8} BG & 5.1 & 0.368 & 0.484 & 0.540 & 0.659 & 0.753 & 0.789 & 0.869 \\
CH & 4.8 & 0.410 & 0.522 & 0.575 & 0.691 & 0.785 & 0.819 & 0.891 \\
\rowcolor[gray]{0.8} CZ & 6.6 & 0.439 & 0.548 & 0.600 & 0.713 & 0.802 & 0.836 & 0.911 \\
DE & 54.2 & 0.454 & 0.551 & 0.601 & 0.716 & 0.810 & 0.849 & 0.934 \\
\rowcolor[gray]{0.8} DK & 3.9 & 0.474 & 0.569 & 0.617 & 0.732 & 0.829 & 0.867 & 0.951 \\
EE & 1.5 & 0.570 & 0.671 & 0.717 & 0.813 & 0.895 & 0.931 & 1.000 \\
\rowcolor[gray]{0.8} ES & 24.3 & 0.461 & 0.556 & 0.600 & 0.697 & 0.794 & 0.837 & 0.933 \\
FI & 9.0 & 0.531 & 0.638 & 0.689 & 0.796 & 0.878 & 0.912 & 0.991 \\
\rowcolor[gray]{0.8} FR & 51.1 & 0.511 & 0.610 & 0.657 & 0.755 & 0.845 & 0.886 & 0.979 \\
GB & 38.5 & 0.547 & 0.641 & 0.687 & 0.787 & 0.873 & 0.911 & 0.996 \\
\rowcolor[gray]{0.8} GR & 5.8 & 0.369 & 0.479 & 0.530 & 0.642 & 0.739 & 0.781 & 0.871 \\
HR & 1.6 & 0.351 & 0.464 & 0.520 & 0.640 & 0.734 & 0.768 & 0.840 \\
\rowcolor[gray]{0.8} HU & 4.4 & 0.390 & 0.496 & 0.548 & 0.663 & 0.753 & 0.786 & 0.860 \\
IE & 3.2 & 0.514 & 0.606 & 0.651 & 0.754 & 0.843 & 0.880 & 0.962 \\
\rowcolor[gray]{0.8} IT & 34.5 & 0.390 & 0.492 & 0.540 & 0.647 & 0.744 & 0.786 & 0.877 \\
LT & 1.5 & 0.538 & 0.640 & 0.688 & 0.789 & 0.874 & 0.912 & 0.997 \\
\rowcolor[gray]{0.8} LU & 0.7 & 0.421 & 0.533 & 0.588 & 0.707 & 0.801 & 0.835 & 0.906 \\
LV & 0.7 & 0.575 & 0.672 & 0.717 & 0.811 & 0.894 & 0.932 & 1.000 \\
\rowcolor[gray]{0.8} NL & 11.5 & 0.451 & 0.546 & 0.596 & 0.716 & 0.813 & 0.851 & 0.933 \\
NO & 13.7 & 0.614 & 0.710 & 0.755 & 0.849 & 0.926 & 0.961 & 1.000 \\
\rowcolor[gray]{0.8} PL & 15.2 & 0.472 & 0.577 & 0.629 & 0.740 & 0.830 & 0.868 & 0.952 \\
PT & 4.8 & 0.412 & 0.511 & 0.559 & 0.661 & 0.756 & 0.797 & 0.887 \\
\rowcolor[gray]{0.8} RO & 5.4 & 0.423 & 0.532 & 0.583 & 0.689 & 0.777 & 0.816 & 0.904 \\
RS & 3.9 & 0.422 & 0.530 & 0.582 & 0.695 & 0.783 & 0.815 & 0.887 \\
\rowcolor[gray]{0.8} SE & 16.6 & 0.572 & 0.672 & 0.719 & 0.818 & 0.901 & 0.937 & 1.000 \\
SI & 1.4 & 0.372 & 0.482 & 0.535 & 0.651 & 0.743 & 0.778 & 0.851 \\
\rowcolor[gray]{0.8} SK & 3.1 & 0.438 & 0.546 & 0.597 & 0.707 & 0.792 & 0.825 & 0.901 \\
Eur. (agg.) & 345.3  & 0.611 & 0.698 & 0.738 & 0.822 & 0.907 & 0.945 & 1.000 \\
\rowcolor[gray]{0.8} Eur. (avg.) & 345.3  & 0.475 & 0.575 & 0.623 & 0.730 & 0.821 & 0.859 & 0.942
\end{longtable}
\end{center}

\begin{center}
\begin{longtable}{l||rrrrrrrr|rrrr}
\caption{
Logistic growth of VRES installation, for the base scenario. Shown are the
penetrations of wind and solar PV production in the total electricity
production, for the reference years 2015-2050 in five-year steps, and the fit
parameters, see Eq.\ \eqref{eq:logfit}. The last two rows contain the average
development in Europe.
}\\
\label{tab:logfit}
{\bf country} & \multicolumn{8}{c|}{\bf penetration in electricity production} &
\multicolumn{4}{c}{\bf fit parameters} \\
& 2015 & 2020 & 2025 & 2030 & 2035 & 2040 & 2045 & 2050 & \multicolumn{1}{c}{$a$} & \multicolumn{1}{c}{$b$} & \multicolumn{1}{c}{$m$} & \multicolumn{1}{c}{$y_0$} \\
\hline
\endfirsthead
{\bf country} & \multicolumn{8}{c|}{\bf penetration in electricity production} &
\multicolumn{4}{c}{\bf fit parameters} \\
& 2015 & 2020 & 2025 & 2030 & 2035 & 2040 & 2045 & 2050 & \multicolumn{1}{c}{$a$} & \multicolumn{1}{c}{$b$} & \multicolumn{1}{c}{$m$} & \multicolumn{1}{c}{$y_0$} \\
\hline
\endhead
AT (wind) & 0.04 & 0.09 & 0.17 & 0.30 & 0.44 & 0.55 & 0.62 & 0.66 & 2.1e-05 & 0.69 & 0.16 & 1967 \\
\nopagebreak
AT (PV) & 0.00 & 0.00 & 0.04 & 0.12 & 0.21 & 0.29 & 0.32 & 0.33 & 1.1e-11 & 0.33 & 0.50 & 1980 \\
\rowcolor[gray]{0.8} BA (wind) & 0.00 & 0.04 & 0.20 & 0.38 & 0.55 & 0.67 & 0.68 & 0.68 & 2.5e-09 & 0.68 & 0.56 & 1990 \\
\nopagebreak
\rowcolor[gray]{0.8} BA (PV) & 0.00 & 0.02 & 0.08 & 0.15 & 0.23 & 0.30 & 0.31 & 0.32 & 4.7e-09 & 0.32 & 0.34 & 1975 \\
BE (wind) & 0.03 & 0.10 & 0.23 & 0.40 & 0.56 & 0.65 & 0.69 & 0.70 & 5.7e-07 & 0.70 & 0.22 & 1965 \\
\nopagebreak
BE (PV) & 0.00 & 0.01 & 0.04 & 0.12 & 0.19 & 0.26 & 0.29 & 0.30 & 3.8e-08 & 0.30 & 0.30 & 1978 \\
\rowcolor[gray]{0.8} BG (wind) & 0.03 & 0.07 & 0.18 & 0.34 & 0.50 & 0.60 & 0.64 & 0.66 & 1.1e-05 & 0.66 & 0.22 & 1979 \\
\nopagebreak
\rowcolor[gray]{0.8} BG (PV) & 0.00 & 0.01 & 0.08 & 0.16 & 0.25 & 0.32 & 0.34 & 0.34 & 1.1e-10 & 0.34 & 0.47 & 1980 \\
CH (wind) & 0.00 & 0.01 & 0.12 & 0.30 & 0.47 & 0.64 & 0.69 & 0.69 & 4.2e-13 & 0.69 & 0.61 & 1980 \\
\nopagebreak
CH (PV) & 0.00 & 0.01 & 0.06 & 0.13 & 0.21 & 0.28 & 0.31 & 0.31 & 6.3e-10 & 0.31 & 0.41 & 1979 \\
\rowcolor[gray]{0.8} CZ (wind) & 0.00 & 0.02 & 0.10 & 0.27 & 0.45 & 0.63 & 0.70 & 0.71 & 1.3e-08 & 0.71 & 0.35 & 1979 \\
\nopagebreak
\rowcolor[gray]{0.8} CZ (PV) & 0.01 & 0.02 & 0.06 & 0.13 & 0.20 & 0.26 & 0.28 & 0.29 & 1.7e-06 & 0.29 & 0.24 & 1979 \\
DE (wind) & 0.13 & 0.21 & 0.31 & 0.43 & 0.53 & 0.61 & 0.66 & 0.69 & 7.5e-05 & 0.73 & 0.13 & 1954 \\
\nopagebreak
DE (PV) & 0.03 & 0.07 & 0.14 & 0.21 & 0.25 & 0.27 & 0.28 & 0.28 & 3.5e-05 & 0.28 & 0.20 & 1979 \\
\rowcolor[gray]{0.8} DK (wind) & 0.38 & 0.49 & 0.58 & 0.64 & 0.68 & 0.71 & 0.72 & 0.73 & 1.8e-02 & 0.74 & 0.12 & 1984 \\
\nopagebreak
\rowcolor[gray]{0.8} DK (PV) & 0.00 & 0.00 & 0.01 & 0.06 & 0.13 & 0.20 & 0.26 & 0.27 & 4.0e-14 & 0.27 & 0.58 & 1980 \\
EE (wind) & 0.02 & 0.05 & 0.13 & 0.28 & 0.47 & 0.63 & 0.71 & 0.75 & 5.0e-06 & 0.77 & 0.21 & 1974 \\
\nopagebreak
EE (PV) & 0.00 & 0.00 & 0.01 & 0.06 & 0.12 & 0.18 & 0.23 & 0.24 & 4.7e-13 & 0.24 & 0.53 & 1980 \\
\rowcolor[gray]{0.8} ES (wind) & 0.18 & 0.27 & 0.36 & 0.46 & 0.54 & 0.61 & 0.65 & 0.67 & 1.0e-03 & 0.71 & 0.11 & 1966 \\
\nopagebreak
\rowcolor[gray]{0.8} ES (PV) & 0.04 & 0.08 & 0.15 & 0.22 & 0.27 & 0.29 & 0.30 & 0.30 & 2.0e-04 & 0.30 & 0.20 & 1988 \\
FI (wind) & 0.01 & 0.06 & 0.20 & 0.40 & 0.60 & 0.74 & 0.78 & 0.79 & 5.5e-06 & 0.80 & 0.29 & 1987 \\
\nopagebreak
FI (PV) & 0.00 & 0.00 & 0.01 & 0.05 & 0.10 & 0.15 & 0.20 & 0.20 & 2.2e-13 & 0.20 & 0.54 & 1980 \\
\rowcolor[gray]{0.8} FR (wind) & 0.04 & 0.10 & 0.23 & 0.42 & 0.59 & 0.69 & 0.73 & 0.75 & 6.0e-05 & 0.76 & 0.21 & 1983 \\
\nopagebreak
\rowcolor[gray]{0.8} FR (PV) & 0.00 & 0.01 & 0.05 & 0.11 & 0.18 & 0.23 & 0.24 & 0.24 & 1.9e-08 & 0.25 & 0.34 & 1980 \\
GB (wind) & 0.08 & 0.21 & 0.40 & 0.59 & 0.71 & 0.76 & 0.78 & 0.79 & 7.9e-05 & 0.79 & 0.22 & 1982 \\
\nopagebreak
GB (PV) & 0.00 & 0.01 & 0.05 & 0.10 & 0.15 & 0.20 & 0.21 & 0.21 & 8.6e-12 & 0.21 & 0.51 & 1980 \\
\rowcolor[gray]{0.8} GR (wind) & 0.13 & 0.25 & 0.39 & 0.51 & 0.59 & 0.62 & 0.63 & 0.64 & 4.1e-05 & 0.64 & 0.19 & 1970 \\
\nopagebreak
\rowcolor[gray]{0.8} GR (PV) & 0.01 & 0.05 & 0.14 & 0.23 & 0.32 & 0.35 & 0.36 & 0.36 & 4.9e-08 & 0.36 & 0.36 & 1980 \\
HR (wind) & 0.01 & 0.04 & 0.14 & 0.29 & 0.45 & 0.58 & 0.62 & 0.64 & 1.0e-05 & 0.64 & 0.27 & 1988 \\
\nopagebreak
HR (PV) & 0.00 & 0.02 & 0.10 & 0.19 & 0.28 & 0.35 & 0.36 & 0.36 & 1.3e-12 & 0.36 & 0.61 & 1981 \\
\rowcolor[gray]{0.8} HU (wind) & 0.01 & 0.03 & 0.11 & 0.26 & 0.43 & 0.57 & 0.64 & 0.66 & 4.6e-07 & 0.67 & 0.26 & 1977 \\
\nopagebreak
\rowcolor[gray]{0.8} HU (PV) & 0.00 & 0.00 & 0.05 & 0.14 & 0.22 & 0.30 & 0.34 & 0.34 & 9.1e-18 & 0.34 & 0.82 & 1980 \\
IE (wind) & 0.14 & 0.24 & 0.35 & 0.48 & 0.58 & 0.66 & 0.71 & 0.74 & 1.6e-04 & 0.77 & 0.13 & 1960 \\
\nopagebreak
IE (PV) & 0.00 & 0.00 & 0.01 & 0.06 & 0.12 & 0.18 & 0.24 & 0.25 & 1.5e-14 & 0.25 & 0.60 & 1980 \\
\rowcolor[gray]{0.8} IT (wind) & 0.03 & 0.07 & 0.15 & 0.28 & 0.43 & 0.55 & 0.61 & 0.64 & 1.6e-05 & 0.66 & 0.19 & 1975 \\
\nopagebreak
\rowcolor[gray]{0.8} IT (PV) & 0.01 & 0.03 & 0.09 & 0.18 & 0.27 & 0.33 & 0.35 & 0.35 & 2.2e-06 & 0.35 & 0.26 & 1983 \\
LT (wind) & 0.04 & 0.09 & 0.21 & 0.39 & 0.57 & 0.68 & 0.73 & 0.75 & 6.8e-06 & 0.76 & 0.20 & 1972 \\
\nopagebreak
LT (PV) & 0.00 & 0.00 & 0.04 & 0.10 & 0.16 & 0.22 & 0.24 & 0.24 & 4.9e-18 & 0.24 & 0.83 & 1980 \\
\rowcolor[gray]{0.8} LU (wind) & 0.01 & 0.04 & 0.12 & 0.28 & 0.46 & 0.61 & 0.68 & 0.70 & 1.5e-06 & 0.71 & 0.24 & 1977 \\
\nopagebreak
\rowcolor[gray]{0.8} LU (PV) & 0.00 & 0.01 & 0.05 & 0.11 & 0.19 & 0.25 & 0.28 & 0.29 & 3.1e-07 & 0.29 & 0.26 & 1978 \\
LV (wind) & 0.04 & 0.11 & 0.25 & 0.44 & 0.62 & 0.71 & 0.74 & 0.75 & 6.9e-05 & 0.76 & 0.23 & 1987 \\
\nopagebreak
LV (PV) & 0.00 & 0.00 & 0.03 & 0.09 & 0.15 & 0.21 & 0.24 & 0.24 & 8.4e-19 & 0.24 & 0.85 & 1980 \\
\rowcolor[gray]{0.8} NL (wind) & 0.11 & 0.24 & 0.40 & 0.55 & 0.64 & 0.69 & 0.71 & 0.71 & 8.3e-04 & 0.72 & 0.19 & 1988 \\
\nopagebreak
\rowcolor[gray]{0.8} NL (PV) & 0.00 & 0.00 & 0.01 & 0.07 & 0.14 & 0.21 & 0.27 & 0.28 & 2.0e-13 & 0.28 & 0.55 & 1980 \\
NO (wind) & 0.05 & 0.19 & 0.40 & 0.61 & 0.78 & 0.83 & 0.85 & 0.85 & 2.1e-05 & 0.85 & 0.30 & 1988 \\
\nopagebreak
NO (PV) & 0.00 & 0.00 & 0.00 & 0.04 & 0.07 & 0.11 & 0.14 & 0.15 & 1.2e-13 & 0.15 & 0.54 & 1980 \\
\rowcolor[gray]{0.8} PL (wind) & 0.03 & 0.09 & 0.23 & 0.42 & 0.59 & 0.69 & 0.72 & 0.74 & 6.2e-05 & 0.74 & 0.24 & 1988 \\
\nopagebreak
\rowcolor[gray]{0.8} PL (PV) & 0.00 & 0.00 & 0.01 & 0.06 & 0.13 & 0.19 & 0.25 & 0.26 & 1.2e-11 & 0.26 & 0.46 & 1980 \\
PT (wind) & 0.17 & 0.27 & 0.38 & 0.48 & 0.56 & 0.61 & 0.64 & 0.65 & 1.4e-04 & 0.67 & 0.13 & 1959 \\
\nopagebreak
PT (PV) & 0.01 & 0.04 & 0.11 & 0.20 & 0.28 & 0.32 & 0.34 & 0.34 & 7.3e-07 & 0.34 & 0.28 & 1980 \\
\rowcolor[gray]{0.8} RO (wind) & 0.02 & 0.11 & 0.29 & 0.46 & 0.62 & 0.68 & 0.69 & 0.69 & 9.5e-06 & 0.69 & 0.36 & 1993 \\
\nopagebreak
\rowcolor[gray]{0.8} RO (PV) & 0.00 & 0.00 & 0.06 & 0.14 & 0.22 & 0.29 & 0.31 & 0.31 & 3.9e-15 & 0.31 & 0.70 & 1980 \\
RS (wind) & 0.00 & 0.04 & 0.21 & 0.38 & 0.56 & 0.68 & 0.69 & 0.69 & 1.7e-08 & 0.70 & 0.57 & 1994 \\
\nopagebreak
RS (PV) & 0.00 & 0.02 & 0.09 & 0.17 & 0.24 & 0.30 & 0.30 & 0.30 & 2.9e-11 & 0.31 & 0.52 & 1981 \\
\rowcolor[gray]{0.8} SE (wind) & 0.04 & 0.09 & 0.20 & 0.38 & 0.58 & 0.71 & 0.78 & 0.81 & 2.3e-05 & 0.83 & 0.20 & 1977 \\
\nopagebreak
\rowcolor[gray]{0.8} SE (PV) & 0.00 & 0.00 & 0.01 & 0.04 & 0.09 & 0.13 & 0.17 & 0.18 & 1.3e-13 & 0.18 & 0.55 & 1980 \\
SI (wind) & 0.00 & 0.01 & 0.14 & 0.30 & 0.47 & 0.62 & 0.65 & 0.65 & 5.4e-14 & 0.65 & 0.67 & 1980 \\
\nopagebreak
SI (PV) & 0.00 & 0.01 & 0.06 & 0.15 & 0.24 & 0.32 & 0.35 & 0.35 & 1.7e-10 & 0.35 & 0.45 & 1980 \\
\rowcolor[gray]{0.8} SK (wind) & 0.00 & 0.02 & 0.13 & 0.31 & 0.49 & 0.66 & 0.70 & 0.71 & 1.6e-10 & 0.71 & 0.47 & 1980 \\
\nopagebreak
\rowcolor[gray]{0.8} SK (PV) & 0.00 & 0.01 & 0.07 & 0.15 & 0.22 & 0.28 & 0.29 & 0.29 & 1.5e-13 & 0.29 & 0.63 & 1980 \\
Avg. (wind) & 0.07 & 0.15 & 0.27 & 0.43 & 0.57 & 0.66 & 0.70 & 0.72 & - & - & - & - \\
\nopagebreak
Avg. (PV) & 0.01 & 0.03 & 0.07 & 0.13 & 0.20 & 0.24 & 0.27 & 0.27 & - & - & - & -
\end{longtable}
\end{center}

\begin{center}
\begin{longtable}{l||r|rrrr|rrrr|rrrr}
\caption{
Line capacities for the reference year-dependent transmission capacity layouts
3, 4, and 5 (see Secs.\ \ref{sec:trnsfixed}--\ref{sec:trns7090}) according to the base scenario in the years 2020, 2030,
2040, and 2050. Net transfer capacities are given in MW.
}\\
\label{tab:linecap}

{\bf link} & {\bf today} & \multicolumn{4}{c|}{\bf 70\,\% benefit} 
 & \multicolumn{4}{c|}{\bf 90\,\% benefit} & \multicolumn{4}{c}{\bf
   unconstrained} \\
 & & 2020 & 2030 & 2040 & 2050 & 2020 & 2030 & 2040 & 2050 & 2020 & 2030 & 2040 & 2050 \\
\hline
\endfirsthead
 {\bf link} & {\bf today} & \multicolumn{4}{c|}{\bf 70\,\% benefit} 
 & \multicolumn{4}{c|}{\bf 90\,\% benefit} & \multicolumn{4}{c}{\bf
   unconstrained} \\
 & & 2020 & 2030 & 2040 & 2050 & 2020 & 2030 & 2040 & 2050 & 2020 & 2030 & 2040 & 2050 \\
\hline
\endhead
  & 470 & 470 & 1210 & 1890 & 2030 & 470 & 2440 & 3460 & 3680 & 470 & 4900 & 12270 & 13720 \\
\nopagebreak 
 \multirow{-2}{*}{AT $\rightleftarrows$ CH} & 1200 & 1200 & 1350 & 1990 & 2100 & 1200 & 2430 & 3310 & 3510 & 1200 & 6250 & 10020 & 11050 \\
\rowcolor[gray]{0.8} & 600 & 600 & 1580 & 2630 & 2850 & 600 & 3480 & 5000 & 5350 & 600 & 8660 & 15500 & 16100 \\
\nopagebreak 
\rowcolor[gray]{0.8} \multirow{-2}{*}{AT $\rightleftarrows$ CZ} & 1000 & 1000 & 1500 & 2530 & 2740 & 1000 & 3280 & 4570 & 4850 & 1000 & 5650 & 10480 & 11720 \\
 & 2000 & 2000 & 2370 & 3940 & 4280 & 2000 & 5300 & 7830 & 8450 & 2000 & 11150 & 25300 & 28170 \\
\nopagebreak 
 \multirow{-2}{*}{AT $\rightleftarrows$ DE} & 2200 & 2200 & 2630 & 4180 & 4460 & 2200 & 5320 & 7260 & 7670 & 2200 & 10520 & 17420 & 17420 \\
\rowcolor[gray]{0.8} & 800 & 800 & 1450 & 2240 & 2410 & 800 & 2850 & 3900 & 4130 & 800 & 6930 & 8910 & 9590 \\
\nopagebreak 
\rowcolor[gray]{0.8} \multirow{-2}{*}{AT $\rightleftarrows$ HU} & 800 & 800 & 1310 & 2220 & 2380 & 800 & 2980 & 4690 & 5120 & 800 & 7340 & 19300 & 19860 \\
 & 220 & 220 & 2200 & 3390 & 3600 & 220 & 4250 & 5710 & 6050 & 220 & 7830 & 13840 & 15460 \\
\nopagebreak 
 \multirow{-2}{*}{AT $\rightleftarrows$ IT} & 285 & 290 & 2190 & 3560 & 3820 & 290 & 4590 & 6560 & 6990 & 290 & 10850 & 19050 & 21450 \\
\rowcolor[gray]{0.8} & 900 & 900 & 1240 & 2000 & 2160 & 900 & 2570 & 3410 & 3600 & 900 & 4360 & 7400 & 7830 \\
\nopagebreak 
\rowcolor[gray]{0.8} \multirow{-2}{*}{AT $\rightleftarrows$ SI} & 900 & 900 & 1180 & 1950 & 2140 & 900 & 2600 & 3910 & 4220 & 900 & 7000 & 14000 & 15500 \\
 & 600 & 600 & 600 & 950 & 1040 & 600 & 1300 & 2020 & 2210 & 600 & 3920 & 7100 & 7290 \\
\nopagebreak 
 \multirow{-2}{*}{BA $\rightleftarrows$ HR} & 600 & 600 & 620 & 930 & 990 & 600 & 1180 & 1590 & 1670 & 600 & 2580 & 3780 & 3850 \\
\rowcolor[gray]{0.8} & 350 & 350 & 420 & 730 & 790 & 350 & 980 & 1460 & 1580 & 350 & 2070 & 4170 & 4230 \\
\nopagebreak 
\rowcolor[gray]{0.8} \multirow{-2}{*}{BA $\rightleftarrows$ RS} & 450 & 450 & 450 & 640 & 680 & 450 & 810 & 1140 & 1210 & 450 & 2620 & 3850 & 4020 \\
 & 550 & 550 & 1040 & 1720 & 1840 & 550 & 2270 & 3410 & 3660 & 550 & 6080 & 15540 & 15780 \\
\nopagebreak 
 \multirow{-2}{*}{BG $\rightleftarrows$ GR} & 500 & 570 & 1180 & 1800 & 1900 & 1270 & 2240 & 3060 & 3270 & 2320 & 7290 & 10280 & 10560 \\
\rowcolor[gray]{0.8} & 600 & 600 & 600 & 940 & 1030 & 600 & 1300 & 1960 & 2110 & 600 & 4800 & 7220 & 7220 \\
\nopagebreak 
\rowcolor[gray]{0.8} \multirow{-2}{*}{BG $\rightleftarrows$ RO} & 600 & 600 & 710 & 1060 & 1130 & 600 & 1300 & 1740 & 1830 & 600 & 3050 & 4820 & 5260 \\
 & 450 & 450 & 790 & 1410 & 1540 & 450 & 1960 & 3010 & 3250 & 450 & 6330 & 10800 & 11230 \\
\nopagebreak 
 \multirow{-2}{*}{BG $\rightleftarrows$ RS} & 300 & 300 & 1080 & 1540 & 1620 & 300 & 1840 & 2340 & 2450 & 300 & 3420 & 5350 & 5790 \\
\rowcolor[gray]{0.8} & 3500 & 3500 & 3500 & 3810 & 4070 & 3500 & 4910 & 7100 & 7620 & 3500 & 9990 & 20700 & 22060 \\
\nopagebreak 
\rowcolor[gray]{0.8} \multirow{-2}{*}{CH $\rightleftarrows$ DE} & 1500 & 1500 & 2450 & 3860 & 4120 & 1500 & 4890 & 6760 & 7210 & 1500 & 9450 & 14730 & 16360 \\
 & 4165 & 4170 & 4170 & 4170 & 4170 & 4170 & 4890 & 6530 & 6860 & 4170 & 8660 & 13850 & 14950 \\
\nopagebreak 
 \multirow{-2}{*}{CH $\rightleftarrows$ IT} & 1810 & 1810 & 2310 & 3790 & 4100 & 1810 & 5000 & 7360 & 7890 & 1810 & 10380 & 23390 & 26300 \\
\rowcolor[gray]{0.8} & 2300 & 2300 & 2300 & 2300 & 2340 & 2300 & 2840 & 4210 & 4580 & 2300 & 5620 & 14760 & 16820 \\
\nopagebreak 
\rowcolor[gray]{0.8} \multirow{-2}{*}{CZ $\rightleftarrows$ DE} & 800 & 800 & 1580 & 2460 & 2620 & 800 & 3070 & 4180 & 4470 & 800 & 8370 & 10870 & 11810 \\
 & 2200 & 2200 & 2200 & 2200 & 2200 & 2200 & 2200 & 2200 & 2340 & 2200 & 3530 & 5030 & 5360 \\
\nopagebreak 
 \multirow{-2}{*}{CZ $\rightleftarrows$ SK} & 1200 & 1200 & 1200 & 1310 & 1420 & 1200 & 1710 & 2530 & 2740 & 1200 & 3710 & 8200 & 8790 \\
\rowcolor[gray]{0.8} & 1550 & 1550 & 1550 & 2250 & 2400 & 1550 & 2900 & 4050 & 4310 & 1550 & 9210 & 11590 & 12130 \\
\nopagebreak 
\rowcolor[gray]{0.8} \multirow{-2}{*}{DE $\rightleftarrows$ DK} & 2085 & 2090 & 2090 & 2430 & 2610 & 2090 & 3180 & 4500 & 4830 & 2090 & 5350 & 10470 & 12920 \\
 & 980 & 980 & 980 & 980 & 980 & 980 & 980 & 980 & 980 & 980 & 980 & 980 & 980 \\
\nopagebreak 
 \multirow{-2}{*}{DE $\rightleftarrows$ LU} & NRL\footnotemark & 980 & 980 & 980 & 980 & 980 & 980 & 980 & 980 & 980 & 980 & 1360 & 1500 \\
\rowcolor[gray]{0.8} & 600 & 600 & 2310 & 3630 & 3860 & 600 & 4620 & 6370 & 6760 & 600 & 13830 & 16430 & 16580 \\
\nopagebreak 
\rowcolor[gray]{0.8} \multirow{-2}{*}{DE $\rightleftarrows$ SE} & 610 & 610 & 2240 & 3710 & 4030 & 610 & 4930 & 7090 & 7580 & 610 & 5020 & 16420 & 20370 \\
 & 750 & 750 & 750 & 750 & 750 & 750 & 750 & 980 & 1040 & 750 & 1730 & 2500 & 2580 \\
\nopagebreak 
 \multirow{-2}{*}{EE $\rightleftarrows$ LV} & 850 & 850 & 850 & 850 & 850 & 850 & 850 & 1060 & 1120 & 850 & 1120 & 2650 & 2910 \\
\rowcolor[gray]{0.8} & 350 & 350 & 490 & 800 & 860 & 350 & 1060 & 1560 & 1680 & 350 & 3000 & 4230 & 4350 \\
\nopagebreak 
\rowcolor[gray]{0.8} \multirow{-2}{*}{FI $\rightleftarrows$ EE} & 350 & 350 & 510 & 810 & 870 & 350 & 990 & 1520 & 1660 & 350 & 990 & 4230 & 5350 \\
 & 1650 & 1650 & 2650 & 4800 & 5240 & 1650 & 6570 & 9650 & 10390 & 1650 & 8280 & 22690 & 25290 \\
\nopagebreak 
 \multirow{-2}{*}{FI $\rightleftarrows$ SE} & 2050 & 2050 & 3100 & 4790 & 5100 & 2050 & 5990 & 7770 & 8170 & 2050 & 12130 & 15070 & 15420 \\
\rowcolor[gray]{0.8} & 3400 & 3400 & 3400 & 3400 & 3400 & 3400 & 3730 & 5310 & 5720 & 3400 & 11000 & 17130 & 18820 \\
\nopagebreak 
\rowcolor[gray]{0.8} \multirow{-2}{*}{FR $\rightleftarrows$ BE} & 2300 & 2300 & 2300 & 3010 & 3200 & 2300 & 3810 & 5360 & 5770 & 2300 & 6650 & 13060 & 13900 \\
 & 3200 & 3200 & 3200 & 3790 & 4090 & 3200 & 4800 & 6720 & 7120 & 3200 & 13990 & 21760 & 25250 \\
\nopagebreak 
 \multirow{-2}{*}{FR $\rightleftarrows$ CH} & 1100 & 1100 & 2330 & 3550 & 3810 & 1100 & 4470 & 6140 & 6550 & 1100 & 10280 & 22430 & 25370 \\
\rowcolor[gray]{0.8} & 2700 & 2700 & 3670 & 5810 & 6230 & 2700 & 7480 & 10590 & 11280 & 2700 & 20680 & 33290 & 35450 \\
\nopagebreak 
\rowcolor[gray]{0.8} \multirow{-2}{*}{FR $\rightleftarrows$ DE} & 3200 & 3200 & 3380 & 5400 & 5810 & 3200 & 7000 & 9850 & 10560 & 3200 & 14430 & 29190 & 29650 \\
 & 1300 & 1300 & 5210 & 8000 & 8530 & 1300 & 10120 & 14040 & 14950 & 1300 & 17440 & 37250 & 37860 \\
\nopagebreak 
 \multirow{-2}{*}{FR $\rightleftarrows$ ES} & 500 & 730 & 3490 & 6550 & 7280 & 3780 & 9310 & 15420 & 16880 & 15070 & 49600 & 65360 & 72750 \\
\rowcolor[gray]{0.8} & 2000 & 2000 & 3320 & 5450 & 5820 & 2000 & 6950 & 9650 & 10270 & 2000 & 20400 & 30070 & 31470 \\
\nopagebreak 
\rowcolor[gray]{0.8} \multirow{-2}{*}{FR $\rightleftarrows$ GB} & 2000 & 2000 & 4150 & 6520 & 6970 & 2000 & 8350 & 11280 & 11940 & 2000 & 16610 & 25930 & 27960 \\
 & 2575 & 2580 & 4180 & 6440 & 6860 & 2580 & 8070 & 10900 & 11530 & 2580 & 20930 & 31670 & 35380 \\
\nopagebreak 
 \multirow{-2}{*}{FR $\rightleftarrows$ IT} & 995 & 1000 & 3750 & 5990 & 6380 & 1000 & 7820 & 11570 & 12450 & 1000 & 16010 & 36580 & 39900 \\
\rowcolor[gray]{0.8} & 450 & 450 & 620 & 1040 & 1130 & 450 & 1340 & 1830 & 1930 & 450 & 3230 & 4510 & 4510 \\
\nopagebreak 
\rowcolor[gray]{0.8} \multirow{-2}{*}{GB $\rightleftarrows$ IE} & 80 & 80 & 700 & 1070 & 1120 & 80 & 1240 & 1680 & 1750 & 80 & 1240 & 3290 & 4230 \\
 & 500 & 630 & 1680 & 2900 & 3110 & 1810 & 3910 & 5930 & 6440 & 2320 & 14070 & 21510 & 21670 \\
\nopagebreak 
 \multirow{-2}{*}{GR $\rightleftarrows$ IT} & 500 & 500 & 2140 & 3100 & 3290 & 500 & 3780 & 4790 & 5010 & 500 & 8200 & 9880 & 11900 \\
\rowcolor[gray]{0.8} & 800 & 800 & 800 & 1160 & 1240 & 800 & 1520 & 2200 & 2370 & 800 & 4150 & 7200 & 7310 \\
\nopagebreak 
\rowcolor[gray]{0.8} \multirow{-2}{*}{HR $\rightleftarrows$ HU} & 1200 & 1200 & 1200 & 1200 & 1200 & 1200 & 1420 & 1910 & 2010 & 1200 & 2240 & 4260 & 4770 \\
 & 350 & 350 & 680 & 990 & 1050 & 350 & 1210 & 1560 & 1630 & 350 & 2480 & 3200 & 3230 \\
\nopagebreak 
 \multirow{-2}{*}{HR $\rightleftarrows$ RS} & 450 & 450 & 530 & 900 & 980 & 450 & 1210 & 1900 & 2040 & 450 & 3910 & 7780 & 7920 \\
\rowcolor[gray]{0.8} & 1000 & 1000 & 1000 & 1520 & 1620 & 1000 & 2010 & 3140 & 3370 & 1000 & 5260 & 12290 & 13030 \\
\nopagebreak 
\rowcolor[gray]{0.8} \multirow{-2}{*}{HR $\rightleftarrows$ SI} & 1000 & 1000 & 1000 & 1570 & 1690 & 1000 & 2030 & 2680 & 2830 & 1000 & 5170 & 6670 & 7100 \\
 & 600 & 600 & 1000 & 1600 & 1710 & 600 & 2090 & 2870 & 3050 & 600 & 3740 & 6330 & 6580 \\
\nopagebreak 
 \multirow{-2}{*}{HU $\rightleftarrows$ RS} & 700 & 700 & 930 & 1590 & 1700 & 700 & 2120 & 3320 & 3640 & 700 & 7110 & 12760 & 13280 \\
\rowcolor[gray]{0.8} & 600 & 600 & 1750 & 3100 & 3350 & 600 & 4150 & 6290 & 6840 & 600 & 12030 & 23040 & 24970 \\
\nopagebreak 
\rowcolor[gray]{0.8} \multirow{-2}{*}{HU $\rightleftarrows$ SK} & 1300 & 1300 & 1810 & 3080 & 3310 & 1300 & 4030 & 5550 & 5920 & 1300 & 6870 & 12890 & 14150 \\
 & 160 & 160 & 1250 & 1930 & 2060 & 160 & 2450 & 3400 & 3620 & 160 & 5550 & 11410 & 12280 \\
\nopagebreak 
 \multirow{-2}{*}{IT $\rightleftarrows$ SI} & 580 & 580 & 1220 & 1810 & 1910 & 580 & 2240 & 3110 & 3340 & 580 & 5030 & 12570 & 12890 \\
\rowcolor[gray]{0.8} & 1300 & 1300 & 1300 & 1300 & 1300 & 1300 & 1300 & 1300 & 1300 & 1300 & 1300 & 1680 & 1720 \\
\nopagebreak 
\rowcolor[gray]{0.8} \multirow{-2}{*}{LV $\rightleftarrows$ LT} & 1500 & 1500 & 1500 & 1500 & 1500 & 1500 & 1500 & 1500 & 1500 & 1500 & 1500 & 1840 & 2190 \\
 & 2400 & 2400 & 2400 & 2570 & 2720 & 2400 & 3280 & 4620 & 4920 & 2400 & 7210 & 12100 & 13080 \\
\nopagebreak 
 \multirow{-2}{*}{NL $\rightleftarrows$ BE} & 2400 & 2400 & 2400 & 2560 & 2750 & 2400 & 3390 & 4860 & 5210 & 2400 & 9970 & 14340 & 15980 \\
\rowcolor[gray]{0.8} & 3000 & 3000 & 3000 & 3270 & 3540 & 3000 & 4260 & 5860 & 6240 & 3000 & 10310 & 17240 & 19010 \\
\nopagebreak 
\rowcolor[gray]{0.8} \multirow{-2}{*}{NL $\rightleftarrows$ DE} & 3850 & 3850 & 3850 & 3850 & 3850 & 3850 & 3850 & 5170 & 5550 & 3850 & 10590 & 16370 & 19280 \\
 & 1000 & 1000 & 2340 & 4020 & 4350 & 1000 & 5270 & 7450 & 7930 & 1000 & 11700 & 24870 & 27920 \\
\nopagebreak 
 \multirow{-2}{*}{NL $\rightleftarrows$ GB} & 1000 & 1000 & 3120 & 5060 & 5400 & 1000 & 6450 & 8780 & 9270 & 1000 & 15420 & 24320 & 25300 \\
\rowcolor[gray]{0.8} & 700 & 700 & 2670 & 3950 & 4180 & 700 & 4870 & 6770 & 7190 & 700 & 14710 & 18290 & 19640 \\
\nopagebreak 
\rowcolor[gray]{0.8} \multirow{-2}{*}{NL $\rightleftarrows$ NO} & 700 & 700 & 2470 & 3930 & 4170 & 700 & 4940 & 6800 & 7210 & 700 & 9720 & 17990 & 20250 \\
 & 950 & 950 & 1350 & 2000 & 2110 & 950 & 2420 & 3090 & 3230 & 950 & 4310 & 7530 & 8120 \\
\nopagebreak 
 \multirow{-2}{*}{NO $\rightleftarrows$ DK} & 950 & 950 & 1190 & 1680 & 1770 & 950 & 2030 & 2600 & 2730 & 950 & 5630 & 6850 & 7510 \\
\rowcolor[gray]{0.8} & 3595 & 3600 & 3600 & 3600 & 3600 & 3600 & 3600 & 3810 & 4020 & 3600 & 9100 & 12180 & 12240 \\
\nopagebreak 
\rowcolor[gray]{0.8} \multirow{-2}{*}{NO $\rightleftarrows$ SE} & 3895 & 3900 & 3900 & 3900 & 3900 & 3900 & 3900 & 3900 & 3900 & 3900 & 5500 & 11950 & 13160 \\
 & 1800 & 1800 & 1800 & 1990 & 2110 & 1800 & 2510 & 3450 & 3680 & 1800 & 3490 & 8010 & 8880 \\
\nopagebreak 
 \multirow{-2}{*}{PL $\rightleftarrows$ CZ} & 800 & 800 & 1170 & 1880 & 2020 & 800 & 2410 & 3370 & 3580 & 800 & 5810 & 10610 & 11570 \\
\rowcolor[gray]{0.8} & 1100 & 1100 & 1510 & 2450 & 2630 & 1100 & 3170 & 4340 & 4630 & 1100 & 4040 & 11470 & 12780 \\
\nopagebreak 
\rowcolor[gray]{0.8} \multirow{-2}{*}{PL $\rightleftarrows$ DE} & 1200 & 1200 & 1630 & 2540 & 2730 & 1200 & 3250 & 4380 & 4640 & 1200 & 8500 & 11260 & 11960 \\
 & 1 & 0 & 1900 & 3000 & 3180 & 0 & 3760 & 5220 & 5580 & 0 & 8630 & 17090 & 18120 \\
\nopagebreak 
 \multirow{-2}{*}{PL $\rightleftarrows$ SE} & 600 & 600 & 1920 & 3140 & 3410 & 600 & 4080 & 5900 & 6270 & 600 & 7680 & 15850 & 16860 \\
\rowcolor[gray]{0.8} & 600 & 600 & 1730 & 2630 & 2790 & 600 & 3280 & 4520 & 4790 & 600 & 5550 & 10930 & 11430 \\
\nopagebreak 
\rowcolor[gray]{0.8} \multirow{-2}{*}{PL $\rightleftarrows$ SK} & 500 & 500 & 1530 & 2520 & 2730 & 500 & 3380 & 4940 & 5350 & 500 & 7990 & 16890 & 17940 \\
 & 1500 & 1500 & 1500 & 2040 & 2190 & 1500 & 2650 & 3840 & 4110 & 2250 & 7720 & 11780 & 12150 \\
\nopagebreak 
 \multirow{-2}{*}{PT $\rightleftarrows$ ES} & 1700 & 1700 & 1700 & 1900 & 2040 & 1700 & 2430 & 3240 & 3400 & 2540 & 5370 & 7850 & 7850 \\
\rowcolor[gray]{0.8} & 700 & 700 & 1240 & 2160 & 2350 & 700 & 2910 & 4470 & 4870 & 700 & 9250 & 15870 & 16330 \\
\nopagebreak 
\rowcolor[gray]{0.8} \multirow{-2}{*}{RO $\rightleftarrows$ HU} & 700 & 700 & 1400 & 2100 & 2240 & 700 & 2630 & 3490 & 3670 & 700 & 4640 & 7160 & 7570 \\
 & 700 & 700 & 700 & 910 & 1000 & 700 & 1250 & 1940 & 2120 & 700 & 3650 & 5480 & 5680 \\
\nopagebreak 
 \multirow{-2}{*}{RO $\rightleftarrows$ RS} & 500 & 500 & 660 & 900 & 950 & 500 & 1080 & 1390 & 1470 & 500 & 2750 & 3830 & 3830 \\
\rowcolor[gray]{0.8} & 1980 & 1980 & 1980 & 1980 & 1980 & 1980 & 2170 & 3060 & 3240 & 1980 & 2210 & 7040 & 8230 \\
\nopagebreak 
\rowcolor[gray]{0.8} \multirow{-2}{*}{SE $\rightleftarrows$ DK} & 2440 & 2440 & 2440 & 2440 & 2440 & 2440 & 2440 & 2900 & 3070 & 2440 & 6590 & 7340 & 7340 \\
\footnotetext{no realistic limit}
\end{longtable}
\end{center}

\end{small}

\end{appendix}

\end{document}